\documentclass[a4paper,12pt]{article} 

\usepackage{graphicx}

\usepackage{a4wide}
\usepackage{amsmath,amssymb}

\usepackage{cite}
\makeatletter
\renewcommand\section{\@startsection {section}{1}{\z@}%
                                   {-3.5ex \@plus -1ex \@minus -.2ex}%
                                   {2.3ex \@plus.2ex}%
                                   {\large\bf}}   
\renewcommand\subsection{\@startsection {subsection}{1}{\z@}%
                                   {-3.5ex \@plus -1ex \@minus -.2ex}%
                                   {2.3ex \@plus.2ex}%
                                   {\normalfont\bf}}         
\makeatother

\allowdisplaybreaks

\DeclareMathOperator{\tr}{tr}
\newcommand{\ii}{\mathrm{i}}

\newcommand{\sbraket}[1]{\lbrack #1\rbrack}
\makeatletter
\newcommand{\fmslash}[2][0mu]{%
  \mathchoice
    {\fmsl@sh\displaystyle{#1}{#2}}%
    {\fmsl@sh\textstyle{#1}{#2}}%
    {\fmsl@sh\scriptstyle{#1}{#2}}%
    {\fmsl@sh\scriptscriptstyle{#1}{#2}}}
\newcommand{\fmsl@sh}[3]{%
  \m@th\ooalign{$\hfil#1\mkern#2/\hfil$\crcr$#1#3$}}
\makeatother

\def\bra#1{\mathinner{\langle{#1}|}}
\def\ket#1{\mathinner{|{#1}\rangle}}
\def\braket#1{\mathinner{\langle{#1}\rangle}}

\newcommand{\greensfunc}[1]{\braket{0|\mathop{\rm T}\left[#1\right]|0}}

\allowdisplaybreaks
\numberwithin{equation}{section}
\numberwithin{table}{section}
\numberwithin{figure}{section}

\begin{document}

\thispagestyle{empty}
\begin{flushright}
{\small
PITHA~08/22 \\
SFB/CPP-08-66\\
arXiv:0809.1442[hep-ph]\\ 
\hfill\\
September 8, 2008
}
\end{flushright}

\vspace{\baselineskip}

\begin{center}
{\bf \large  Twistor-inspired construction of massive quark amplitudes}
\end{center}

\begin{center}
 {\sc Christian Schwinn}\footnote{schwinn@physik.rwth-aachen.de}\\\hfill\\
{\normalsize \it  Institut f\"ur Theoretische Physik E}\\ 
{\normalsize \it RWTH Aachen University, D - 52056 Aachen, Germany}
\end{center}

\date{}

\setcounter{page}{0}
\begin{abstract}
  The analog of the Cachazo~-~Svr\v{c}ek~-~Witten rules for scattering
  amplitudes with massive quarks is derived following an approach
  previously employed for amplitudes with massive scalars. A
  prescription for the external wave-functions is given that leads to
  a one-to-one relation between fields in the action and spin-states
  of massive quarks.  Several examples for the application of the
  rules are given and the structure of some all-multiplicity
  amplitudes with a pair of massive quarks is discussed.  The rules
  make supersymmetric relations to amplitudes with massive scalars
  manifest at the level of the action.  The formalism is extended to
  several quark flavors with different masses.

\end{abstract}
\newpage

\section{Introduction}
In the intriguing diagrammatic construction of Cachazo~-~Svr\v{c}ek~-~Witten
(CSW)~\cite{Cachazo:2004kj} tree level QCD amplitudes
are constructed from vertices that are off-shell continuations of
maximal helicity violating (MHV) amplitudes
with two negative helicity gluons and an arbitrary number of positive helicity
gluons~\cite{Parke:1986gb,Berends:1987me},
 \begin{equation}
\label{eq:mhv}
  A_n(g_1^+,\dots,g_i^-,\dots, g_j^-,\dots g_n^+)
=\ii 2^{n/2-1}
\frac{\braket{ij}^4}{\braket{12}\braket{23}\dots\braket{(n-1)n}\braket{n1}}
\end{equation}
where plus and minus labels are connected
by scalar propagators. This formalism has been extended to
massless
quarks~\cite{Wu:2004fb} and  supersymmetric 
theories~\cite{Georgiou:2004wu}.  
While recursive methods~\cite{Berends:1987me,Britto:2005fq} tend to lead to
more compact analytical expressions and better numerical
 performance~\cite{Dinsdale:2006sq},
the CSW rules have been  useful in several respects such as
one-loop calculations
 in supersymmetric~\cite{Brandhuber:2004yw} and  non-supersymmetric
gauge theories~\cite{Bedford:2004nh,Ettle:2007qc} and the investigation of
multiple collinear and soft limits in QCD~\cite{Birthwright:2005ak}.

Since the CSW rules were originally motivated by arguments from 
twistor-string theory~\cite{Witten:2003nn}, they were  thought 
to be limited mainly to massless particles, although extensions
 to currents with single color-neutral
external massive particles, i.e. 
electroweak gauge bosons~\cite{Bern:2004ba} or Higgs
bosons~\cite{Dixon:2004za}, have been found.
However, both for phenomenological and theoretical reasons it is interesting
to explore how far these rules can be extended to massive, colored
particles, for instance massive quarks such as the top quark. 

Meanwhile several field theoretical derivations of the CSW rules independent of
twistor strings have been 
given~\cite{Britto:2005fq,Risager:2005vk,Gorsky:2005sf,Mansfield:2005yd,Ettle:2006bw,Boels:2007qn}.
Particular interesting for the extension towards massive particles are derivations of the CSW rules based on field redefinitions in the framework of 
 light-cone gauge Yang-Mills~\cite{Mansfield:2005yd} or the lift of the Yang-Mills action
to twistor space~\cite{Mason:2005zm,Boels:2007qn}.
Both of these action-based methods  have been used to construct CSW rules for a
 propagating massive, colored  scalar $\phi$~\cite{Boels:2007pj,Boels:2008ef}.
 In this derivation the same field redefinition that results in
the CSW rules 
for massless scalars was applied for a massive scalar. As a result
 the transformation of the scalar mass-term  gives rise to a tower of vertices
with only positive helicity gluons
\begin{equation}
\label{eq:csw-mass}
V_{\text{CSW}}(\bar \phi_1,g^+_2,\dots \phi_n)= \ii  2^{n/2-1}
\frac{- m^2 \braket{1n}}{
\braket{12}\dots\braket{(n-1)n }} 
\end{equation}
while the remaining vertices are identical to those of  massless
scalars~\cite{Georgiou:2004wu}. 
Interestingly, the vertex~\eqref{eq:csw-mass} involves only `holomorphic' spinor-products and therefore allows a twistor-space interpretation of scattering amplitudes with massive scalars~\cite{Boels:2007pj} that is not obvious from the
explicit results of on-shell scattering amplitudes~\cite{Badger:2005zh}.
The fact
that the new vertex~\eqref{eq:csw-mass} for massive scalars
 is proportional to the mass  suggests the
possibility of constructing a systematic expansion of scattering
amplitudes in the mass. The leading term was discussed
in~\cite{Boels:2007pj}.  Furthermore, the CSW rules have been used~\cite{Boels:2007pj} to
simplify the proof of  on-shell recursion relations for massive
scalars~\cite{Badger:2005zh,Schwinn:2007ee}.

In this paper the approach of~\cite{Boels:2007pj,Boels:2008ef} is extended to massive quarks, using the same field redefinitions obtained  recently 
in the derivation of CSW rules for  massless quarks~\cite{Ettle:2008ey,Boels:2008du}. 
 The CSW vertices for massive quarks 
 derived in this paper are related to the massive scalar 
vertex~\eqref{eq:csw-mass}
by the same relations implied by the application of supersymmetric Ward identities~(SUSY-WIs)~\cite{Grisaru:1976vm}
to scattering amplitudes with massive quarks and scalars~\cite{Schwinn:2006ca}.
A similar approach to CSW rules for massive quarks was proposed in~\cite{Ettle:2008ey} where a complication related to the external polarization spinors of massive quarks was encountered\footnote{While this paper was in  preparation, a revised version of~\cite{Ettle:2008ey} appeared which also 
states the all-multiplicity vertices for massive quarks. However, neither a detailed derivation nor non-trivial examples beyond the three point amplitude are provided.}. As a result, scattering amplitudes for massive
quarks with a given helicity had to be obtained by adding different
amplitudes of the fields appearing in the Lagrangian. In the present paper it
is shown how to avoid these complications by an appropriate choice of the
quantization axis of the massive-quark spin. The rules
are also shown to generalize to amplitudes
with several quark flavors of different masses.

The organization of this paper is as follows.  As the canonical
transformation method proposed in~\cite{Mansfield:2005yd} is based on
the light-cone gauge, section~\ref{sec:light-cone} reviews the
light-cone treatment of massive quarks, slightly generalizing the
treatment given in~\cite{Mansfield:2005yd,Ettle:2008ey}. It is shown
that an appropriate choice of external spinors can be used to arrive
at a one-to-one relation between the dynamical degrees of freedom
present in the light-cone gauge Lagrangian and the external
spin-states, avoiding the complication encountered
in~\cite{Ettle:2008ey}. The resulting Feynman rules are shown to be
equivalent to those obtained earlier in a diagrammatic
approach~\cite{Schwinn:2005pi}.  Section~\ref{sec:derive} derives the
CSW vertices for massive quarks and discusses some features of the
resulting diagrammatics, after summarizing the field
redefinitions~\cite{Ettle:2006bw,Ettle:2008ey,Boels:2008du} utilized
in the derivation.  Some examples for the application of the rules are
given in section~\ref{sec:applications}, together with a discussion of
the structure of all-multiplicity amplitudes with a pair of massive
quarks and an arbitrary number of positive helicity gluons and one
negative helicity gluon. Finally the extension of the rules to
amplitudes with several quark-flavors of different masses is
discussed.  The notation used for the color decomposition of
scattering amplitudes and the MHV amplitudes for quarks, the SUSY-WIs
relating amplitudes of massive quarks to those of massive scalars, and
some technical details of the derivation of the vertices are contained
in appendices.

\section{Light-cone QCD and scalar diagrams with massive quarks}
\label{sec:light-cone}
The  method for the derivation of the  CSW rules introduced
in~\cite{Mansfield:2005yd,Ettle:2006bw} and applied to quarks 
in~\cite{Ettle:2008ey} uses the light-cone approach to
Yang-Mills theory. In this section the treatment of massive quarks in
light-cone Yang-Mills is reviewed  and streamlined compared
to~\cite{Ettle:2008ey}. The resulting Feynman rules are shown 
to agree with those obtained  previously from a diagrammatic analysis~\cite{Schwinn:2005pi}.
\subsection{Light-cone QCD}
The starting point is the Lagrangian of Yang-Mills theory
coupled to a massive quark 
\begin{equation}
\label{eq:qcd}
\mathcal{L}_{\text{QCD}}=  -\frac{1}{2}\tr[F^{\mu\nu}F_{\mu\nu}]+\bar\Psi(\ii\fmslash
D-m)\Psi
\end{equation}
where $D_\mu=\partial_\mu-\ii g A_\mu$ and $A_\mu=A^a_\mu T^a$ with $T^a$ the
generators in the fundamental representation of $SU(N)$. The light-cone gauge
is obtained by fixing a light-like vector $n_+$ and imposing the condition
$(n_+\cdot A)=0$.
To implement this condition it is useful to decompose
 four-momenta and vector fields  into light-cone components according to
\begin{equation}
  p_\mu=p_+ n_{-,\mu} + p_- n_{+,\mu}+ p_{\perp,\mu} 
\end{equation}
with  two light-like vectors $n_+,n_-$ normalized according to $(n_-\cdot
n_+)=1$ and with $p_\perp\cdot n_+=p_\perp\cdot n_-=0$. 
Introducing light-like basis vectors $n_{z/\bar z}$ for the transverse components 
satisfying $(n_z\cdot n_{\bar z})=-1$ one can further decompose 
\begin{equation}
  p_{\perp,\mu}=p_z n_{\bar z,\mu}+p_{\bar z} n_{z,\mu}
\end{equation}
so that $p_\perp\cdot k_\perp=-p_zk_{\bar z}-p_{\bar z}k_z $.
For the moment it is not necessary to specify 
the basis-vectors $n_\pm$ and $n_{z/\bar z}$  further. 

To incorporate massive quarks it is convenient to follow closely the setup
used in the
construction of soft-collinear effective
theory~\cite{Bauer:2000yr} for massive
quarks~\cite{Rothstein:2003wh}.
 The treatment of
quarks in~\cite{Mansfield:2005yd,Ettle:2008ey} emerges as a special case for
the choice $n_{\pm}=2^{-1/2}(1,0,0,\pm 1)$.
A Dirac spinor  $\Psi$ and it's conjugate are 
decomposed as 
\begin{equation}
\label{eq:def-chi}
  \Psi=\frac{\fmslash n_-\fmslash n_+}{2}\chi
  +\frac{\fmslash n_+\fmslash n_-}{2}\zeta \;,\quad
 \bar \Psi=\bar \chi\frac{\fmslash n_+\fmslash n_-}{2}
  +\bar \zeta\frac{\fmslash n_-\fmslash n_+}{2}.
\end{equation}
In the light-cone gauge $(n_+\cdot A)=A_+=0$ the Lagrangian~\eqref{eq:qcd} becomes
\begin{multline}
\mathcal{L}
=\tr\left[(\partial_+A_-)^2-2\partial_
    + A_\perp\left(\partial_-A_\perp-\partial_\perp A_--\ii g
      [A_-,A_\perp]\right)-F_{z\bar z}^2\right]\\
+\ii\bar\chi  \fmslash n_+ D_-\chi +\ii \bar\zeta \fmslash n_- \partial_+\zeta
-m(\bar\chi\zeta+\bar\zeta\chi)+\ii\bar\chi\fmslash D_\perp\zeta
+\ii\bar\zeta\fmslash D_\perp\chi \, .
\end{multline}  
In light-cone quantization $\partial_-$ is treated as a time
derivative.
Since there is no term with a $\partial_-$ derivative acting on the 
 fields $A_-$ and $\zeta$,
 they are considered as
non-dynamical and 
 can be eliminated from the Lagrangian using their equations of motion
\begin{align}
\label{eq:eom}
    A_{-,ij}&=\frac{1}{\partial_+^2}
\left(-[D_\perp,\partial_+A_{\perp,ij}]
+ \frac{g}{2}\left(\bar\chi_j \fmslash n_+\chi_i 
-\frac{1}{N}\delta_{ij}(\bar\chi\fmslash n_+\chi) \right)\right)\\
\zeta&=\frac{1}{2}\frac{1}{\ii \partial_+}(\ii\fmslash D_\perp +m)
\fmslash n_+\chi\,.
\end{align}
Here the identity 
\begin{equation}
  \frac{\delta A_{ij}(x)}{\delta A_{kl}(y)}=\frac{1}{2}
\left(\delta_{il}\delta_{jk}-\frac{1}{N}\delta_{ij}\delta_{kl}\right)
\delta^4(x-y)
\end{equation}
has been used.
To simplify some of the resulting expressions we introduce a notation
for the product of  vectors or
matrices in color space with the $1/N$ contribution
subtracted:
\begin{equation}
\psi_k \chi_l
\left(\delta_{ik}\delta_{jl}-\frac{1}{N}\delta_{ij}\delta_{kl}\right)
\equiv (\psi\otimes \chi)_{ij}\;,\quad
 \tr[ AB]-\frac{1}{N}\tr[A]\tr[B]
\equiv A\otimes B\,.
\end{equation}

Inserting the solutions of the equations of motion, 
the resulting Lagrangian for the physical degrees of freedom $A_\perp$ and 
$\chi$ has the form
\begin{equation}
\begin{aligned}
\label{eq:light-cone-lag}
\mathcal{L}=\mathcal{L}_{A_\perp}&+\bar\chi \left[\ii \partial_-+ 
(\ii \fmslash D_\perp-m)\frac{1}{2\ii\partial_+}(\ii \fmslash D_\perp+m)
+g \left(\frac{1}{\partial_+^2}[D_\perp,\partial_+A_{\perp}]\right)
\right]\fmslash n_+\chi \\
&+\left(\frac{g}{2}\right)^2(\bar\chi \fmslash n_+\chi)\otimes
\frac{1}{\partial_+^2}(\bar\chi\fmslash n_+\chi)\, .
\end{aligned}
\end{equation}
Note that the derivatives in the last term in the square bracket
do not act on the fermion field to the right.
 The gluon Lagrangian $\mathcal{L}_{A_\perp}$ can be grouped into several terms
according to the field content~\cite{Mansfield:2005yd}:
\begin{equation}
\label{eq:A-lag}
\mathcal{L}_{A_\perp}=
\mathcal{L}^{(2)}_{A_zA_{\bar z}}+\mathcal{L}^{(3)}_{A_zA_zA_{\bar z}}
+\mathcal{L}^{(3)}_{A_zA_{\bar z}A_{\bar z}}
+\mathcal{L}^{(4)}_{A_zA_zA_{\bar z}A_{\bar z}}\,.
\end{equation}
The explicit form will not be needed in the following. 

\subsection{Two-component form of the Lagrangian}
The light-cone formalism is closely related to the
spinor-helicity formalism conventionally used in the CSW
rules~\cite{Mansfield:2005yd} (see also~\cite{Chalmers:1998jb}).
Following the notation  
used in~\cite{Boels:2008ef}, this relation can be established by
 introducing a basis
$(\eta^{\dot \alpha},\kappa^{\dot \alpha})$ of holomorphic spinors and an
anti-holomorphic basis $(\eta^{\alpha},\kappa^{\alpha})$ normalized according
to $\braket{\eta\kappa}=\sbraket{\kappa\eta}=\sqrt 2$, where the brackets are
defined as 
\begin{equation}
\braket{ p q } =  p^{\dot \alpha} q_{\dot\alpha},\quad
[ q p ] = q_{\alpha} p^\alpha\, .
\end{equation}
Here the same conventions for raising and lowering indices
as in~\cite{Boels:2008ef} are used.
Translating from a
 four-vector to
a two-component spinor notation via the mapping $n^{\alpha\dot\beta}=n_\mu
\bar\sigma^{\mu \alpha \dot\beta}$ the basis  vectors in the light-cone
decomposition can be chosen in terms of the spinor-basis as 
\begin{equation}
\label{eq:def-n}
  n_+^{\alpha\dot\alpha}=\eta^{\alpha}\eta^{\dot\alpha},\quad 
  n_-^{\alpha\dot\alpha}=\kappa^{\alpha}\kappa^{\dot\alpha},\quad
  n_{\bar z}^{\alpha\dot\alpha}=\kappa^{\alpha}\eta^{\dot\alpha},\quad
  n_{z}^{\alpha\dot\alpha}=\eta^{\alpha}\kappa^{\dot\alpha}\, .
\end{equation}
The two-component spinors  associated to a momentum $p$ can be expanded in the $(\eta,\kappa)$ bases
as
\begin{equation}
\label{eq:define-spinors}
  p^\alpha=p_+^{-1/2} 
(p_{\bar z}\,\eta^{\alpha}+p_+\,\kappa^{\alpha})
\;,\quad
p^{\dot\alpha}=p_+^{-1/2} 
(p_{z}\,\eta^{\dot\alpha}+p_+\,\kappa^{\dot\alpha}) \, ,
\end{equation}
up to an arbitrary phase. For negative or complex $p_+$ the square
root should be interpreted as $p_+^{1/2} =|p_+|^{1/2}e^{\ii\phi_p/2}$
with the phase defined by $p_+=|p_+|e^{\ii\phi_+}$.
 The expressions of spinor-products in
terms of the light-cone components are given by
\begin{equation}
\label{eq:lc-brakets}
  \begin{aligned}
    \braket{pk}&=
2^{3/2}  \frac{ (p_z k_+-k_zp_+)}{\sbraket{\eta p}\sbraket{\eta k}},&&
    \sbraket{kp}&= 
    2^{3/2}  \frac{(k_{\bar z}p_+-p_{\bar z} k_+)}{
      \braket{\eta p}\braket{\eta k}} \, .
   \end{aligned}
\end{equation}
These expressions can also be defined for off-shell (or massive ) momenta
since they are independent of the $-$ components of the momenta.
This corresponds
to the usual off-shell continuation in the CSW rules where spinors $p^\alpha$ and $p^{\dot\alpha}$ can be associated to an off-shell momentum $p$ by the 
decomposition
\begin{equation}\label{eq:continue}
p^{\dot \alpha} p^{\alpha}=p^{\dot\alpha \alpha}
-\frac{p^2}{2(p\cdot \eta)}\eta^{\dot\alpha}\eta^{\alpha}\,.
\end{equation}
 
The Dirac spinors used up to now are translated 
to a two-component notation by writing
\begin{equation}
\label{eq:weyl}
  \Psi=
  \begin{pmatrix}
    \Psi_{+,\dot\alpha} \\ \bar \Psi_-^\alpha
  \end{pmatrix},\quad
\bar\Psi=
\begin{pmatrix}
\Psi_-^{\dot\alpha},&\bar \Psi_{+,\dot\alpha}
\end{pmatrix}\,.
\end{equation}
In the two-component notation the definition~\eqref{eq:def-chi} of the
light-cone field $\chi$ becomes
\begin{align}
\label{eq:translate-spinors}
  \chi_{+,\dot\alpha}&=
\frac{1}{\sqrt 2}\kappa_{\dot\alpha}(\eta^{\dot\beta}\Psi_{+,\dot\beta})
\equiv \frac{1}{\sqrt 2}\kappa_{\dot\alpha}\chi_+
\quad ,&
\bar \chi_{-}^\alpha&=
-\frac{1}{\sqrt 2}\kappa^{\alpha}(\eta_{\beta}\bar \Psi_{-}^\beta)
\equiv-\frac{1}{\sqrt 2}\kappa^{\alpha}\bar\chi_-
\end{align}
where a short-hand for  the  non-vanishing
components
$  \bar \chi_-=(\eta_\alpha \bar \chi^\alpha_-)$
and $
\chi_+=(\eta^{\dot\alpha}\chi_{+,\dot\alpha})$
has been introduced. For the conjugate spinor the corresponding definitions read
$ \chi_-=(\chi^{\dot\alpha}_-\eta_{\dot\alpha})$
and $
\bar \chi_+=(\bar\chi_{+,\alpha}\eta^{\alpha})$.
In this notation the Lagrangian~\eqref{eq:light-cone-lag} can be rewritten as
\begin{equation}
\begin{aligned}
\label{eq:spinor-lag}
\mathcal{L}_\chi&=
\chi_- \left(\ii\partial_- 
 + D_z\frac{1}{\ii\partial_+}D_{\bar z}-g \frac{1}{\partial_+^2}[D_\perp,\partial_+A_{\perp}] - m^2 \frac{1}{2\ii\partial_+}\right) \bar\chi_- \\
&+ \bar \chi_+ \left(\ii\partial_- + D_{\bar z}\frac{1}{\ii\partial_+}D_{ z}
  - g \frac{1}{\partial_+^2}[D_\perp,\partial_+A_{\perp}] -
m^2 \frac{1}{2\ii\partial_+}\right)\chi_+\\
& +\frac{ g m}{\sqrt 2}\left( 
\chi_- \left[ A_{z},\frac{1}{\ii\partial_+}\right]\chi_+
-\bar\chi_+ \left[A_{\bar
      z},\frac{1}{\ii\partial_+}\right]\bar \chi_-
\right) \\
&+\left(\frac{g}{2}\right)^2(\bar\chi_+\chi_++\chi_-\bar \chi_-)\otimes
\frac{1}{\partial_+^2}(\bar\chi_+\chi_++\chi_-\bar \chi_-) \, .
\end{aligned}
\end{equation}
Note that the first two lines describe two uncoupled massive fields with the  propagator
\begin{equation}
\label{eq:chi-prop}
 \greensfunc{\bar\chi_-(x)\chi_-(y)}= \greensfunc{\chi_+(x)\bar\chi_+(y)}
 =\int\frac{d^4 p}{(2\pi)^4}e^{-\ii p(x-y)}
\frac{ 2\ii p_+}{p^2-m^2}\, .
\end{equation}
The fields are coupled only through the terms in the last two lines,
 in particular
for vanishing mass there is only a quartic coupling. 
\subsection{External states}

In order to calculate scattering amplitudes using the
Lagrangian~\eqref{eq:spinor-lag} the polarization vectors and spinors have to
be specified. Here it is desirable to have a simple relation between the
external polarization states and the fields $A_{z/\bar z}$ and $\chi_\pm$
appearing in the Lagrangian. In particular the external states should be
chosen in such a way that the non-dynamical fields $A_-$ and $\zeta$ that have
been integrated out decouple and do not contribute to correlation functions.
For the polarization vectors of the gluons it is appropriate~\cite{Mansfield:2005yd}
to use the usual expressions of the spinor-helicity
formalism,
\begin{equation}
\label{eq:polarization}
  \epsilon^{+,\alpha\dot\alpha}(k)=\sqrt 2  \,
 \frac{ k^{\alpha} \eta^{\dot\alpha}}{\braket{\eta k}} \quad;\quad
 \epsilon^{-,\alpha\dot\alpha}(k)=\sqrt 2  \,
 \frac{ \eta^{\alpha} k^{\dot\alpha}}{\sbraket{k\eta}}  \, .
\end{equation}
The reference spinors $\eta^\alpha$ and $\eta^{\dot\alpha}$ 
are chosen to be identical for all gluons and are taken to be the same as in
the light-cone decomposition~\eqref{eq:def-n}. A similar gauge has been used
in~\cite{Chalmers:1998jb}.
The  relevant light-cone components of the polarization vectors 
are given by
\begin{equation}
\label{eq:lc-eps}
  \epsilon_{z}^+(k)=\frac{\sbraket{\eta  k}}{\braket{k\eta }}\qquad
  \epsilon_{\bar z}^-(k)=\frac{\braket{\eta k}}{\sbraket{k\eta}} \, .
\end{equation}
Therefore the component $A_z$ of a gauge field $A$ can be identified with 
the positive helicity mode, the component $A_{\bar z}$ with the negative
helicity mode. 
In the conventions used in~\cite{Boels:2008ef} these factors can be
set to one but this will not be assumed in this section.

There are several options for
the definition of the external polarizations of massive quarks 
in the framework of the the spinor-helicity
formalism~\cite{Kleiss:1985yh}, either fixing a reference axis of the quark spin
or using physical helicity state.
A convenient definition of spinors with a spin-axis 
defined  in terms of reference spinors $q$ is given by~\cite{Schwinn:2005pi,Schwinn:2007ee}:
\begin{equation}
\label{eq:massive-spinors}
    v(\pm) = \frac{\left( \fmslash p - m \right) \ket{q \pm}}{\braket{ p\mp|q \pm}}\;,\quad 
\bar{u}(\pm) = \frac{ \bra{q \mp}
 \left( \fmslash p + m \right)}{\braket{q\mp|  p\pm}} \, .
\end{equation}
Here the spinors associated to the massive quark momentum $p$ 
are defined in analogy to the off-shell continuation~\eqref{eq:continue} used in the CSW rules with $\eta$ replaced by $q$.
Since the spinors~\eqref{eq:massive-spinors} have a smooth massless limit,
the considerations in this section apply equally to massless and massive fermions.
Because the quantization axis of the 
quark spin is fixed by the spinors $\ket{q\pm}$~\cite{Kleiss:1985yh},
these reference spinors are not unphysical quantities
that have to drop out in the final result for the helicity amplitude, in contrast to the case of the gluon polarization vectors.
Amplitudes with different  spin-axes can be related as discussed in~\cite{Schwinn:2007ee}.
In an abuse of notation, the massive quark states labeled by plus and minus
labels will sometimes be denoted as positive or negative helicity states, this
should always be understood as referring to the eigenvalue of the spin projectors.

In~\cite{Ettle:2008ey} a  helicity basis for
massive quarks was constructed that 
eliminates the correlation functions of the non-dynamical components $\zeta$
but that does not lead to a one-to-one relation between polarization states and the light-cone
fields so the physical scattering amplitudes have to be assembled from
 different correlation functions of the fields in the Lagrangian.
In order to decouple the non-dynamical quark field $\zeta$ and to 
obtain a one-to-one relationship between the spin-labels and the
light-cone quarks fields $\chi_\pm$ it is
 advantageous to use the quark spinors~\eqref{eq:massive-spinors} and
chose the same reference spinors for all quarks
and set them equal to the spinors $\eta$ used for the light-cone
decomposition.
This can be seen by relating scattering amplitudes to 
 correlation functions of the original fields $\Psi$ using the LSZ formula:
\begin{align}
\label{eq:lsz}
&  \mathcal{A}(\dots\Psi_k(\pm) )=\lim_{k^2\to m^2}\ii
\int d^4 x e^{-\ii k\cdot x}
\greensfunc{\dots \bar\Psi(x)} (-\fmslash k-m) v(k,\pm)\\
&=\lim_{k^2\to m^2}\!\!\frac{(-\ii)(k^2-m^2)}{\braket{k\mp|q\pm}}
\!\!\!\int d^4 x e^{-\ii k\cdot x}
\left(\greensfunc{\dots \bar\chi(x)}\frac{\fmslash n_+\fmslash n_-}{2}
+\greensfunc{\dots\bar\zeta(x)}\frac{\fmslash n_-\fmslash n_+}{2}\right)
\ket{q\pm}\nonumber
\end{align}
For the choice $\ket{q\pm}=\ket{\eta\pm}$ the contribution of
the non-dynamical field $\zeta$ vanishes.
Expressing the
matrix element in terms of an amputated correlation function, one obtains  e.g.:
\begin{equation}
\label{eq:physical-amplitude}
\begin{aligned}
    \mathcal{A}(\dots\Psi_k(+) )
&=\lim_{k^2\to m^2}(-\ii)(k^2-m^2)\int d^4 x e^{-\ii k\cdot x}
\greensfunc{\dots
  \chi_-(x)}\frac{1}{\braket{k\eta }}\\
=&\braket{\dots \bar\chi_-(k))}_{\text{amp}} \sbraket{\eta k} \, .
\end{aligned}
\end{equation}
Here the notation introduced in~\eqref{eq:translate-spinors} was used.
In the last step the amputated correlation function
has been introduced by stripping off
the non-canonically normalized propagator~\eqref{eq:chi-prop}
connecting $\chi_\pm$ and $\bar\chi_\pm$ and using
the identity $2p_+=\braket{\eta p}\sbraket{p\eta}$. 

Treating the other polarizations similarly, one arrives at the following results:
\begin{equation}
\label{eq:physical-amplitudes}
\begin{aligned}
  \mathcal{A}(\dots,\Psi_k(+))
&=\braket{\dots \bar\chi_{-}(k)}_{\text{amp}} \sbraket{\eta k} ,&
    \mathcal{A}(\bar \Psi_k(+),\dots )
&=\braket{\bar\chi_{+}(k)\dots}_{\text{amp}} \sbraket{k\eta} \,,
\\
 \mathcal{A}(\dots,\Psi_k(-) )
&=\braket{\dots \chi_{+}(k)}_{\text{amp}} \braket{\eta k} ,&
  \mathcal{A}(\bar \Psi_k(-),\dots )&=
\braket{\chi_{-}(k)\dots}_{\text{amp}} \braket{k\eta} \, .
\end{aligned}
\end{equation}
Both the external 
normalization factors in~\eqref{eq:physical-amplitudes} and the numerator
factor $2p_+$ in the propagator can be absorbed in the vertices
as discussed below.

\subsection{Scalar diagrams for massive quarks}
\label{sec:lc-vertices}
Using the relation between amplitudes of polarized quarks and correlation
functions of light-cone fields in~\eqref{eq:physical-amplitudes} it is
straightforward to derive  diagrammatic rules from the
Lagrangian~\eqref{eq:spinor-lag}. In order to work with canonically
normalized fields and to make the relation to the physical polarization states
clearer it is convenient to redefine the momentum-space fields as follows:
\begin{equation}
\label{eq:normalize}
  \begin{aligned}
    \bar\chi_-(k)&=Q^+_k\sbraket{\eta k}, &  
    \chi_+(k) &=Q^-_k\braket{\eta k},\\
   \bar\chi_+(k) &=\bar Q^+_k\sbraket{k\eta}, &  
   \chi_-(k) &=\bar Q^-_k\braket{k\eta} \, .
  \end{aligned}
\end{equation}
In these conventions, canonically normalized scalar propagators connect
plus and minus labels:
\begin{equation}
  \greensfunc{Q^{\pm}(x)\bar Q^{\mp}(y)}=\int\frac{d^4 p}{(2\pi)^4}e^{-\ii p(x-y)}
 \frac{\ii}{k^2-m^2}
\end{equation}
It is also seen from~\eqref{eq:physical-amplitude} that the
redefinitions~\eqref{eq:normalize} together with the now canonical propagator
ensure that the external wave-function factors are eliminated.
The vertices of the $Q$ fields are obtained from those of the  $\chi$ fields
simply by multiplying with the appropriate $\braket{\eta\pm|k\mp}$
factors. 
Indicating the field content, the resulting Lagrangian is of the schematic
form
\begin{equation}
  \label{eq:Q-lag}
\mathcal{L}_{Q}=
\mathcal{L}^{(2)}_{QQ}+
\sum_{\lambda=\pm}\left[\mathcal{L}^{(3)}_{\bar Q^\lambda A_zQ^{-\lambda}}
+\mathcal{L}^{(3)}_{\bar Q^\lambda A_{\bar z}Q^{-\lambda}}
+\mathcal{L}^{(4)}_{Q^\lambda A_zA_{\bar
    z}Q^{-\lambda}}\right]
+\mathcal{L}_{Q^4}^{(4)}
+\mathcal{L}_{\text{flip}}
\end{equation}
with 
\begin{equation}
\label{eq:L-flip}
 \mathcal{L}_{\text{flip}}=\mathcal{L}^{(3)}_{\bar Q^- A_{ z}Q^-}
+\mathcal{L}^{(3)}_{\bar Q^+ A_{\bar z}Q^+}  \, .
\end{equation}
The kinetic terms and the
 `helicity conserving' vertices involving opposite labels of the quarks
displayed explicitly in~\eqref{eq:Q-lag}
arise from the first two lines of the Lagrangian~\eqref{eq:spinor-lag}
and the four-quark vertices while the terms proportional to $m$ 
in the third line  give rise to the `helicity flip' vertices with the same
labels of the quarks in~\eqref{eq:L-flip}.

The vertices for the canonically normalized fields
can be read from the Lagrangian~\eqref{eq:spinor-lag}
and translated to the spinor product notation using~\eqref{eq:lc-brakets}.
In order to calculate color-ordered partial amplitudes as summarized in
appendix~\ref{app:color},  the color matrices and coupling constants
 can be stripped off the vertices.
For the cubic vertices one obtains, for instance,
\begin{subequations}
\label{eq:lcg-vertices}
\begin{align}
  V_3(\bar Q^+_1, A_{2,\bar z},Q^-_3)
&=\ii\left(\frac{k_{2,z}}{k_{2,+}}-\frac{k_{3,z}}{k_{3,+}}\right)
\sbraket{1\eta}\braket{\eta 3}\epsilon_{\bar z}^-(k_2)\nonumber\\
&=\sqrt 2 \ii 
\braket{23}\frac{\sbraket{\eta 1}}{\sbraket{\eta 2}}
=\sqrt 2 \ii \frac{\braket{23}^2}{\braket{31}} \, ,
\\
 V_3(\bar Q^+_1, A_{2,\bar z},Q^+_3) &=\frac{\ii  m}{\sqrt 2}
\left(\frac{1}{k_{3,+}}-\frac{1}{k_{2,3,+}}\right)
\sbraket{1\eta}\sbraket{\eta 3}\epsilon_{\bar z}^-(k_2)\nonumber\\
&=(\sqrt 2\ii  m)\frac{\braket{\eta 2}^2}{\braket{\eta 1}\braket{\eta 3}} \, ,
\label{eq:vertex-flip++}\\
 V_3(\bar Q^-_1, A_{2,z},Q^-_3) &=\frac{\ii  m}{\sqrt 2}
\left(\frac{1}{k_{2,3,+}}-\frac{1}{k_{3,+}}\right)
\braket{1\eta}\braket{\eta 3}\epsilon_{z}^+(k_2)\nonumber\\
&=(-\sqrt 2\ii  m)\frac{\sbraket{\eta 2}^2}{\sbraket{\eta 1}\sbraket{\eta 3}}
=(-\sqrt 2\ii  m)\frac{\braket{1 3}^2}{\braket{12}\braket{2 3}} \, .
\label{eq:vertex-flip--}
\end{align}
\end{subequations}
Here all particles are treated as outgoing and the notation
$k_{i,j}=k_i+k_{i+1}+\dots k_j$ was introduced.
 In order to facilitate a later comparison with the CSW rules, the 
vertices have been expressed in terms of only holomorphic spinors.
The
vertices~\eqref{eq:lcg-vertices} agree precisely with the ones derived
in~\cite{Schwinn:2005pi} from a diagrammatical analysis of QCD in an
axial gauge, 
 up to possible phase differences due to a different momentum routing along
the fermion lines (compare the discussion in section~\ref{sec:rules}).

\section{CSW vertices for massive quarks}
\label{sec:derive}
In this section the Lagrangian~\eqref{eq:spinor-lag} is used as
starting point to derive CSW rules for massive quarks.  The
construction follows the approach of~\cite{Mansfield:2005yd} where
non-local canonical transformations are used to eliminate non MHV-type
vertices in favor of infinite towers of MHV vertices.  This method is
briefly reviewed in subsection~\ref{sec:trafos} and the expressions
for the field redefinitions for gluons~\cite{Ettle:2006bw} and
massless quarks~\cite{Ettle:2008ey} are summarized.
The same expressions can also be obtained in the twistor Yang-Mills
approach~\cite{Boels:2008ef,Boels:2008du} where the CSW rules are derived
from an action for gauge theory on twistor space
by fixing a particular gauge choice~\cite{Mason:2005zm,Boels:2007qn}.

The extension of the CSW rules to massive quarks is given in subsection~\ref{sec:vertices}. 
There are in principle several ways to define `CSW rules' for massive quarks:
\begin{itemize}
\item Use the same field redefinitions as for massless quarks and
  insert them into the light-cone Lagrangian of massive
  quarks~\eqref{eq:spinor-lag}, together with the transformation of
  the gluons.  The rules derived in this way will include the same MHV
  vertices as for massless quarks and new vertices proportional to
  $m^2$ or $m$ resulting from the transformation of the mass-terms and
  the `helicity flip' vertices in the light-cone
  Lagrangian~\eqref{eq:spinor-lag}.  This is analogous to the approach
  used to derive the vertex~\eqref{eq:csw-mass} for massive
  scalars~\cite{Boels:2007pj,Boels:2008ef}.
\item Only transform the gluon fields but not the quark fields.  
  In this approach the vertices of MHV-type helicity structure will
  not be given by off-shell continuations of MHV amplitudes, for
  instance the 4-quark vertex in the light-cone Lagrangian will not be
  dressed with positive helicity gluons.  Therefore the massless limit
  does not reproduce the usual CSW rules for massless quarks.

\item Try to find a mass-dependent redefinition that transforms the
  mass term into a quadratic term of the new fields.  However, it is
  unclear that this would lead to a practical formalism since there
  are non-vanishing on-shell amplitudes that are apparently eliminated
  by such a transformation and that would have to be generated in a
  different way (presumably from `equivalence theorem
  violations'~\cite{Ettle:2007qc}). One could also attempt to
  eliminate the helicity flip vertices in~\eqref{eq:Q-lag} but similar
  remarks apply here.
\end{itemize}
 
In this work the first approach will be used to define the extension of the 
CSW rules. All necessary
 new vertices present for massive quarks
are derived in subsection~\ref{sec:vertices}.  As suggested by
the above remarks, it is doubtful that the other approaches would
lead to simpler rules.  Several technical details are relegated to 
appendix~\ref{app:details}. The resulting rules are summarized in~\ref{sec:rules} where also the structure of the resulting tree-diagrams is analyzed.
 The vertices obtained 
using the same method are also stated in the revised version
 of~\cite{Ettle:2008ey}, however there the treatment of external states led to a more complicated formalism for the calculation of scattering amplitudes
and the structure of the diagrams was not discussed in detail.

\subsection{Field redefinitions and the massless CSW Lagrangian}
\label{sec:trafos}

In the light-cone Lagrangian of the transverse gluonic degrees of
freedom~\eqref{eq:A-lag} all terms apart from the term
$\mathcal{L}^{(3)}_{A_zA_zA_{\bar z}}$ have the same helicity structure as the
MHV-amplitudes~\eqref{eq:mhv}. 
Similarly in the light-cone Lagrangian of massless quarks obtained
from~\eqref{eq:Q-lag} 
all the terms apart from the terms $\mathcal{L}^{(3)}_{\bar Q^+ A_zQ^-}$ and 
$\mathcal{L}^{(3)}_{\bar Q^- A_zQ^+}$ have the right helicity combinations
for the CSW vertices (recall that the helicity flip vertices vanish for
massless quarks).
It was therefore proposed
in\cite{Mansfield:2005yd} to derive the CSW rules 
by a transformation to new gluon variables
$B$ and $\bar B$ and quark variables $\psi_\pm$ and $\bar\psi_\pm$
that satisfies the condition
\begin{equation}
\label{eq:transform-L}
\int d^3 x \left[\mathcal{L}^{(2)}_{A_{\bar
      z}A_z}+\mathcal{L}^{(3)}_{A_zA_zA_{\bar z}}
+\mathcal{L}^{(2)}_{\bar Q Q }+\mathcal{L}^{(3)}_{\bar Q^+ A_zQ^-}
+\mathcal{L}^{(3)}_{\bar Q^- A_zQ^+}
\right]
=\int d^3 x\left(\mathcal{L}^{(2)}_{\bar B B}+\mathcal{L}^{(2)}_{\bar \psi\psi}\right) \, .
\end{equation}
The precise form of the transformation is further constrained by
the requirement to have a trivial Jacobian.
 In momentum space, the transformations of $A_z$ and
$A_{\bar z}$ can be taken to have the form
\begin{align}
  A_{p,z}=&\sum_{n=1}^\infty\int_{1\dots n}
\mathcal{Y}(p,k_1,\dots,k_n)B_{-k_1}\dots B_{-k_n} \, ,
\label{eq:b-trafo}\\
A_{p,\bar z}=&\sum_{n=1}^\infty\sum_{s=1}^n  
\int_{1\dots n} 
\mathcal{X}^s(p,k_1,\dots,k_n)
B_{-k_1}\dots B_{-k_{s-1}}
\left(\frac{(k_{s})_+}{p_+}\bar B_{-k_s}\right) B_{-k_{s+1}}
+A_{p,\bar z}|_{\psi\bar\psi} \, .
\label{eq:bbar-trafo}
\end{align}
where a delta-function implementing momentum conservation, 
$(2\pi)^3\delta^3(p+\sum_ik_i)$ is kept implicit.
The integration measure is defined by
\begin{equation}
\int_{1\dots n}=\prod_{i=1}^n\int
\frac{dk_{i+} dk_{iz} dk_{i\bar z}}{(2\pi)^3}.
\end{equation}
The explicit solutions of the gluon  transformation
have been obtained using the canonical approach in~\cite{Ettle:2006bw}.
In the present conventions the solutions read 
\begin{align}
\label{eq:em-coeff}
 \mathcal{Y}(p,k_1,\dots,k_n)&= 
 \frac{ (g \sqrt 2)^{n-1}\braket{\eta p}^2}{\braket{\eta 1}
  \braket{12}\dots\braket{(n-1) n}\braket{n \eta}} \, ,\\
\mathcal{X}^s(p,k_1,\dots,k_n)&=-\frac{k^s_+}{p_+} 
\mathcal{Y}(p,k_1,\dots,k_n)
 =-\frac{ (g \sqrt 2)^{n-1}\braket{\eta s}^2}{\braket{\eta 1}
  \braket{12}\dots\braket{(n-1) n}\braket{n \eta}}\, .
\label{eq:em-coeff-bbar}
\end{align}
 The
coefficients~\eqref{eq:em-coeff} and~\eqref{eq:em-coeff-bbar} have
also been derived from the twistor Yang-Mills
approach~\cite{Boels:2008ef}.
These expressions hold for spinor phase conventions where the external
wave-function factors of the gluons are trivial, i.e. where
$\braket{\eta k}=\sbraket{k\eta}=\sqrt{2 k_0}$.  These conventions will
be adopted in the remainder of this subsection and
in~\ref{sec:vertices}.  The final results for the diagrammatic rules are
independent of this assumption since modifications in the coefficients
of the field redefinitions will be compensated by non-trivial external
wave-function factors~\eqref{eq:lc-eps}.

The additional term in the transformation of $A_{\bar z}$ in~\eqref{eq:bbar-trafo} involving quark 
fields has the form 
\begin{equation}
  \label{eq:bbar-em-quark}
\left.  p_+ (A_{p,\bar z})\right|_{\bar\psi\psi}
=\sum_{n=2}^\infty
\sum_{s=1}^{n-1}\sum_{\lambda=\pm}
 \int_{1\dots n}
\mathcal{K}_\lambda^s(p,k_1,\dots k_n)
\left(B_{-k_{1}}\dots \psi^{-\lambda}_{-k_s}\right)\otimes
\left(\bar\psi^\lambda_{-k_{s+1}}\dots B_{-k_n}\right) \, .
\end{equation}
The explicit expressions for the coefficients  read in our conventions
\begin{equation}
\label{eq:k-coeff}
\begin{aligned}
\mathcal{K}_-^s(p,k_1,\dots,k_n)&   =-\frac{ (g \sqrt 2)^{n-1}\braket{\eta (s+1)}^3\braket{\eta s}}{\braket{\eta 1}
  \braket{12}\dots\braket{(n-1) n}\braket{n \eta}} \, ,\\
\mathcal{K}_+^s(p,k_1,\dots,k_n)&
=\frac{ (g \sqrt 2)^{n-1}\braket{\eta s}^3\braket{\eta (s+1)}}{
\braket{\eta 1}  \braket{12}\dots\braket{(n-1) n}\braket{n \eta}} \, .
\end{aligned}
\end{equation}
These results are obtained from~\cite{Ettle:2008ey}  by  translating 
their expressions to spinor bracket notation using~\eqref{eq:lc-brakets} 
 and by taking the redefinitions~\eqref{eq:normalize}
into account in order to work with trivial external normalization factors
and canonical scalar propagators.

 The Ansatz for the transformations of the
fermions is taken as
\begin{equation}
\label{eq:q-trafo}
Q^\pm_{p} =  \sum_{n=1}^{\infty}\int_{1\dots n}  \Sigma^\pm(p,k_1,\dots,k_n)
  \left(B_{-k_1} \ldots B_{-k_{n-1}} \Psi^\pm_{-k_n}\right) 
\end{equation}
with the coefficients~\cite{Ettle:2008ey,Boels:2008du}\footnote{
Compared to~\cite{Boels:2008du} the role 
of barred and unbarred fields is exchanged in~\eqref{eq:weyl} so 
the formulas  in~\cite{Boels:2008du} translate to the
present notation as  $(\bar\nu^{\dot\alpha},\nu^\alpha)\to
(\chi_-^{\dot\alpha},\bar\chi_-^{\alpha} )$ and
$(\bar\nu,\nu)\to(\bar Q^-,Q^+)$.}
\begin{equation}
\label{eq:sigma}
\begin{aligned}
  \Sigma^+(p,k_1,\dots,k_n)&=-(g\sqrt{2})^{n-1} \frac{\braket{\eta
      p}}{\braket{\eta 1} \braket{1 2} \ldots \braket{(n-1) n}}\, ,\\
\Sigma^-(p,k_1,\dots,k_n)&=(g\sqrt{2})^{n-1} \frac{\braket{\eta n}^2 }{\braket{\eta1}\braket{1 2}  \ldots \braket{(n-1)n}\braket{\eta p}} \, .
\end{aligned}
\end{equation}
These results are consistent with  the proposed transformation
of the light-cone $\mathcal{N}=4$  superfield~\cite{Feng:2006yy}.
The same coefficients enter the transformations
of the conjugate spinors which we write for clarity as
\begin{equation}
  \bar{Q}^\pm_p =  \sum_{n=1}^{\infty} \int_{1\dots n} 
\bar\Sigma^\pm(p,k_1,\dots,k_n)
  \left( \bar{\Psi}^\pm_{-k_1} B_{-k_2} \ldots B_{-k_n}\right)
\end{equation}
with
\begin{equation}
\label{eq:sigmab}
\begin{aligned}
  \bar\Sigma^+(p,k_1,\dots,k_n)&=(g\sqrt{2})^{n-1} \frac{\braket{\eta
      p}}{\braket{1 2} \ldots \braket{(n-1) n}\braket{n\eta}} \, ,\\
\bar \Sigma^-(p,k_1,\dots,k_n)&=-(g\sqrt{2})^{n-1} \frac{\braket{\eta 1}^2 }{\braket{\eta p}\braket{1 2}  \ldots \braket{(n-1)n}\braket{n\eta }} \, .
\end{aligned}
\end{equation}
The coefficients of the transformation 
 for the quarks 
are related in a simple way to the coefficients
$\mathcal{Z}$
in the corresponding transformation of scalars~\cite{Boels:2008ef}:
\begin{equation}
\label{eq:em-susy}
  \Sigma^+(p,k_1,\dots,k_n)
=-\frac{\braket{\eta p}}{\braket{\eta n}}
  \mathcal{Z}(p,k_1,\dots k_n) ,\quad
\Sigma^-(p,k_1,\dots,k_n)
= \frac{\braket{\eta n}}{\braket{\eta p}}
  \mathcal{Z}(p,k_1,\dots k_n)\, .
\end{equation}

In the next step  the field redefinitions given above
are inserted into the massless
light-cone Lagrangian $\mathcal{L}_{A_\perp}+\mathcal{L}_Q$, where the helicity flip vertices in the quark Lagrangian
~\eqref{eq:Q-lag} vanish in the massless limit. 
Using the fact that the redefinitions satisfy the condition~\eqref{eq:transform-L} and conserve the number of negative helicity gluons it is seen that
the new Lagrangian only contains vertices with the helicity content
of MHV amplitudes:
\begin{equation} 
\mathcal{L}_{A_\perp} +\mathcal{L}_Q=\mathcal{L}^{(2)}_{B\bar B}+
\mathcal{L}^{(2)}_{\bar\psi\psi}+
\sum_{n=3}^\infty \left[\mathcal{L}^{(n)}_{\bar B B\dots \bar BB\dots}
+\mathcal{L}^{(n)}_{\bar\psi,B\dots \bar B B\dots \psi}\right] 
+\sum_{n=4}^\infty\mathcal{L}^{(n)}_{\bar\psi B\dots \psi
\bar\psi B\dots \psi}\,.
\end{equation}
The gluon vertices in momentum space take the form
\begin{equation}
\label{eq:csw-lag}
L^{(n)}_{\bar B B\dots \bar B B\dots}=
\int d^3 x \mathcal{L}^{(n)}_{\bar B B\dots \bar B B\dots}=
\frac{1}{2} \sum_{j=2}^n\int_{1\dots n} 
\mathcal{V}_{\bar B_1, B_2,\dots \bar B_j,\dots B_n}
\tr \left( \bar B_{k_1}\dots\bar B_{k_{j}}\dots B_{k_n}\right)\,.
\end{equation}
It has been argued~\cite{Mansfield:2005yd} and explicitly checked up
 to $n=5$~\cite{Ettle:2006bw} that the coefficients $
 \mathcal{V}_{\bar B_1,\dots \bar B_i,\dots \dots B_n}$ are just the
 MHV amplitudes~\eqref{eq:mhv} continued off-shell according to the  prescription~\eqref{eq:continue}.
 Similarly the triple and quartic quark vertices with the MHV helicity
 structure get dressed with infinitely many positive helicity gluons
and are transformed into the appropriate vertices 
\begin{align}
\label{eq:csw-lag-q2}
L^{(n)}_{\bar\psi,B\dots \bar B B\dots \psi}= 
\sum_{\lambda=\pm}\sum_{j=2}^{n-1}
&\int_{1\dots n} 
\mathcal{V}_{\bar \psi^\lambda_1, B_2,\dots \bar B_j,\dots \psi^{-\lambda}_n}
\left( \bar \psi^{\lambda}_{k_1}\dots\bar B_{k_{j}}\dots \psi^{-\lambda}_{k_n}\right)\,,\\
L^{(n)}_{\bar\psi B\dots \psi
\bar\psi B\dots \psi}=
\sum_{\lambda,\sigma=\pm}&\sum_{j=2}^{n-1}
\int_{1\dots n} \left[
\mathcal{V}_{\bar\psi^{\lambda}_1 B_2\dots \psi^{-\sigma}_j
\bar\psi_{j+1}^\sigma B_{j+2}\dots \psi_n^{-\lambda}} 
(\bar\psi^{\lambda}_{k_1} B_{k_2}\dots \psi^{-\sigma}_{k_j})(
\bar\psi_{k_{j+1}}^\sigma B_{k_{j+2}}\dots \psi_{k_n}^{-\lambda})\right.\nonumber\\
+\frac{1}{N}&\left.
\mathcal{V}_{\bar\psi^{\lambda}_1 B_2\dots \psi^{-\lambda}_j
\bar\psi_j^\sigma B\dots \psi_n^{-\sigma}} 
(\bar\psi^{\lambda}_{k_1} B_{k_2}\dots \psi^{-\lambda}_{k_j})(
\bar\psi_{k_{j+1}}^\sigma B_{k_{j+2}}\dots \psi_{k_n}^{-\sigma})\right]\,.
\label{eq:csw-lag-q4}
\end{align}
Because of the the additional quark contributions in the
transformation of $\partial_+A_{\bar z}$ in~\eqref{eq:bbar-em-quark},
the transformations of the vertices are more intricate than in the
pure gluon case, for instance the two-quark MHV vertices receive
contributions from four sources: the transformed quark-gluon vertices
$\mathcal{L}^{(3)}_{\bar Q^\pm A_{\bar z}Q^\mp}$ and
$\mathcal{L}^{(4)}_{Q^\pm A_zA_{\bar z}Q^\mp}$, but also from the cubic
and quartic gluon vertices $\mathcal{L}^{(3)}_{A_zA_zA_{\bar z}}$ and
$\mathcal{L}^{(4)}_{A_zA_zA_{\bar z}A_{\bar z}}$.

One can argue~\cite{Ettle:2008ey} that the vertex functions
in~\eqref{eq:csw-lag-q2} and~\eqref{eq:csw-lag-q4} are indeed 
off-shell continuations of the MHV amplitudes summarized in appendix~\ref{app:color}. This argument is based on the fact that the vertices must reproduce
the on-shell MHV amplitudes, the fact that they are independent of the
$k_-$ components of the momenta, and that singularities in the coefficients
of the field redefinitions that could lead to contributions to on-shell
scattering amplitudes~\cite{Ettle:2007qc} are limited to the three-point vertices (see also~\cite{Boels:2008ef}).
It also has been checked explicitly that the 
four- and five point MHV vertices with a quark pair and the 
four-quark four-point MHV vertex are
obtained  correctly~\cite{Ettle:2008ey}. 
 Additional evidence 
 comes from the fact that the transformations  also
can be obtained in the twistor-Yang-Mills approach~\cite{Boels:2008du}
that has been used earlier to derive the CSW rules including
quarks~\cite{Boels:2007qn}.

The color structures of the vertices~\eqref{eq:csw-lag-q2}
and~\eqref{eq:csw-lag-q4} agrees with that of the corresponding
amplitudes reviewed in appendix~\ref{app:color}.  It is also seen
that color-ordered diagrams computed with these vertices give rise to
the associated color structures. As example 
consider a diagram with a two-quark MHV
vertex~\eqref{eq:csw-lag-q2} and a gluonic MHV vertex~\eqref{eq:csw-lag}
of the structure
\begin{multline}
\mathcal{V}_{\bar \psi_1^+, B_P,B_{j+1}\dots \bar B_l,\dots \psi^{-}_n}
(T^{a_1}T^{a_P}T^{a_{j+1}}\dots T^{a_n})_{i_1 i_n}
\frac{\ii}{P^2}  \mathcal{V}_{\bar B_{P} B_{2}\dots \bar B_{i}\dots B_{j}}
\tr[T^{a_P}T^{a_2}\dots T^{a_j}]  \\
\Rightarrow\left(\mathcal{V}_{\bar \psi_1^+, B_P,B_{j+1}\dots \bar B_l,\dots \psi^{-}_n}
\frac{\ii}{P^2}  \mathcal{V}_{\bar B_{P} B_{2}\dots \bar B_{i}\dots B_{j}}
\right)
(T^{a_1}T^{a_2}\dots T^{a_n})_{i_1 i_n}\,.
\end{multline}
Here the color Fierz-identity~\eqref{eq:fierz} was applied to get the second line. The $1/N$ term in the Fierz identity gives rise to a term proportional 
to the trace $\tr[T^{a_2}\dots T^{a_j}]$ that vanishes since
the gluonic MHV vertices satisfy a $U(1)$ decoupling identity~\cite{Mangano:1990by}.
For four-quark amplitudes both leading and sub-leading contributions are generated:
\begin{multline}
\mathcal{V}_{\bar \psi_1^+, B_{2}\dots B_{i-1},
\bar B_P,B_{j+1},\dots \psi^{-}_n}
(T^{a_1}\dots T^{a_P}\dots T^{a_n})_{i_1 i_n}
\frac{\ii}{P^2}
\mathcal{V}_{\bar\psi_{i+1}^+B_{i+2} \dots \bar B_{j}B_{P}\psi_i^-}
(T^{a_{i+1}}\dots  T^{a_P}T^{a_i})_{i_{i+1},i_i}  \\
\Rightarrow\left(\mathcal{V}_{\bar \psi_1^+, B_{2}\dots B_{i-1},
\bar B_P,B_{j+1},\dots \psi^{-}_n}
\frac{\ii}{P^2}
  \mathcal{V}_{\bar\psi_{i+1}^+B_{i+2} \dots \bar B_{j}B_{P}\psi_i^-}\right)
\Bigl[(T^{a_1}\dots T^{a_i})_{i_1i_i}(T^{a_{i+1}}\dots T^{a_n})_{i_{i+1}i_n}\\
-\frac{1}{N}(T^{a_1}\dots T^{a_{i-1}}T^{a_{j+1}}\dots T^{a_n} )_{i_1i_n}
(T^{a_{i+1}}\dots T^{a_j}T^{a_i})_{i_{i+1}i_i}
\Bigr]\,.
\end{multline}
Therefore this diagram contributes both to the leading 
and sub-leading color-amplitudes
\begin{equation}
A_n(\bar \psi_1,B_2,\dots \psi_i,\bar\psi_{i+1},\dots \bar B_j,\psi_n)
\quad,\quad
B_n(\bar \psi_1,\dots B_{i-1},B_{j+1},\dots \psi_n;\bar\psi_{i+1},\dots \bar B_j,\psi_i)\,.
\end{equation}
 appearing in the color decomposition~\eqref{eq:color-4q}.
This structure is easy to visualize in the double line notation, see~\eqref{eq:color-propagator}. The leading color structure arises from the `$U(N)$-gluon' propagator with one color-line connecting $\bar\psi_1$ to $\psi_i$ and the
other  line connecting  $\bar\psi_{i+1}$ to $\psi_n$. The sub-leading structure
is due to the `$U(1)$-gluon' where the color-flow agrees with the fermion-number flow.
Also the  4-quark MHV vertices~\eqref{eq:csw-lag-q4} contribute to both 
leading and sub-leading color structures.

\subsection{Derivation of CSW vertices for massive quarks}
\label{sec:vertices}

In this subsection the new CSW vertices for massive quarks will be derived.
As outlined in the beginning of this section, following the method used to
derive the CSW rules for a colored massive
scalar~\cite{Boels:2007pj,Boels:2008ef} the same field redefinitions as in the
massless case will be inserted into the massive Lagrangian.  This results in
the same vertices as in the CSW rules for massless quarks and additional
vertices proportional to the mass.  All vertices not present already for
massless quarks will be derived explicitly, leaving technical details to
appendix~\ref{app:details}.

Analogously to  the scalar vertex~\eqref{eq:csw-mass}, vertices 
proportional to $m^2$ and with an arbitrary
number of positive-helicity gluons arise from inserting the field
redefinitions into the mass terms
in the first two lines
of~\eqref{eq:spinor-lag}, e.g.:
\begin{equation}
\int d^3 x \left(-m^2\bar Q^-Q^+\right)= \sum_{n=2}^\infty\int_{1\dots n} 
\mathcal{V}_{\bar \psi^-_1,B_{2},\dots B_{n-1},\psi^+_n}
 \left(\bar\psi^-_{k_1}B_{k_2}\dots B_{k_{n-1}}\psi^+_{k_n}\right)\,.
\end{equation}
The close similarity of the transformations 
of quarks and scalars~\eqref{eq:em-susy} also implies a simple
relation between the vertex functions of fermions and scalars. 
Indeed one finds for the explicit expression of the vertex functions:
\begin{equation}
\label{eq:csw+++}
\mathcal{V}_{\bar \psi^-_1,B_{2},\dots B_{n-1},\psi^+_n}
=-\frac{\braket{\eta 1}}{\braket{\eta n}}
  \mathcal{V}_{\bar \phi_1,B_{2},\dots B_{n-1},\phi_n}=
\frac{- m^2(g\sqrt 2)^{n-2}\braket{1n}\braket{\eta 1} }{
\braket{12}\dots\braket{(n-1)n }\braket{n\eta }}  \,.
\end{equation}
Therefore these vertices manifestly respect the SUSY-WI~\eqref{eq:conserve+++} of the corresponding on-shell amplitudes.
Note that the vertices depend explicitly on the spinor $\ket{\eta+}$. This
reflects the dependence of scattering amplitude on the spin quantization axis.

In addition to the vertices generated by the transformation of the mass term,
also  the helicity flip
vertices  $\mathcal{L}^{(3)}_{\bar Q^- A_zQ^-}$ and 
$\mathcal{L}^{(3)}_{\bar Q^+ A_{\bar z}Q^+}$  give rise to new towers of vertices.
Inserting the field redefinitions, the 
$\bar Q^-A_zQ^-$  helicity flip vertex~\eqref{eq:vertex-flip--} becomes:
\begin{equation}
\int d^3 x\mathcal{L}_{\bar Q^- A_z Q^-}=
\sum_{n=2}^\infty\int_{1\dots n} 
\mathcal{V}_{\bar \psi^-_1,B_{2},\dots B_{n-1},\psi^-_n}
 \left(\bar\psi^-_{k_1}B_{k_2}\dots B_{k_{n-1}}\psi^-_{k_n}\right)\,.
\end{equation}
The vertex function is given in terms of the 
 coefficients~\eqref{eq:em-coeff} and~\eqref{eq:sigma}: 
\begin{equation}
\label{eq:flip--trafo}
\begin{aligned}
  \mathcal{V}_{\bar \psi^-_1,B_{2},\dots,\psi^-_n}
&=\frac{g m}{\sqrt 2}\sum_{i=1}^{n-2}\sum_{j=i+1}^{n-1}
\left[  
\left(\frac{1}{ k_{(i+1,n),+}}-\frac{1}{{ k_{(j+1,n),+}}}\right)
 \braket{k_{1,i}\eta}\braket{\eta k_{j+1,n}}
\right.\\
&\left.\times\bar\Sigma^-(-k_{1,i},k_1,\dots k_i)
\mathcal{Y}(-k_{i+1,j},k_{i+1},\dots k_j)\Sigma^-(-k_{j+1,n},k_{j+1},\dots k_n)
\right]\,.
\end{aligned}
\end{equation}
The somewhat tedious evaluation of the double sum is described in
 appendix~\ref{app:details}.
As a result one obtains the simple expression
\begin{equation}
 \mathcal{V}_{\bar \psi^-_1,B_{2},\dots, \dots B_{n-1},\psi^-_n}
=\frac{ -m (g\sqrt 2)^{n-2} \braket{1n}^2}{
\braket{12}\dots\braket{(n-1)n }} 
=\frac{\braket{1n}}{m}
  \mathcal{V}_{\bar \phi_1,B_{2},\dots B_{n-1},\phi_n}\,.
\end{equation}
As indicated,  this vertex satisfies the same SUSY-WI~\eqref{eq:flip+++}
as the corresponding on-shell amplitude.

The last remaining vertex in the Lagrangian~\eqref{eq:spinor-lag} that
is proportional to the mass is the helicity flip vertex $\bar
Q^+A_{\bar z}Q^+$~\eqref{eq:vertex-flip++}.  As in the previous
case the field redefinitions dress the vertex
with an arbitrary number of positive helicity gluons.  In addition,
however, the quark-contribution in the transformation of $A_{\bar
  z}$~\eqref{eq:bbar-em-quark} leads to additional towers of
four-quark vertices:
\begin{equation}
\label{eq:trafo+-+}
\int d^3 x\mathcal{L}_{\bar Q^+ A_{\bar z} Q^+}
=\sum_{n=3}^\infty
L^{(n)}_{\bar Q^+,B\dots \bar BB\dots Q^+}+
\sum_{n=4}^\infty
L^{(n)}_{\bar Q^+,B\dots Q\bar QB\dots Q^+}\,.
\end{equation}
The two-quark terms are of the form
\begin{equation}
L^{(n)}_{\bar Q^+,B\dots \bar BB\dots Q^+}= \int_{1\dots n} 
\sum_{s=2}^{n-1}
\mathcal{V}_{\bar \psi^+_1,B_{2},\dots \bar B_s,\dots, B_{n-1},\psi^+_n}
 \left(\bar\psi^+_{k_1}B_{k_2}\dots \bar B_{k_s}\dots B_{k_{n-1}}\psi^+_{k_n}
\right)
\end{equation}
where the vertex function is obtained by inserting
the field redefinitions~\eqref{eq:em-coeff-bbar}
and~\eqref{eq:sigma}:
\begin{align}
  \mathcal{V}_{\bar \psi^+_1,B_{2},\dots ,\bar B_s,\dots,\psi^+_n}
&=\frac{g m}{\sqrt 2}\sum_{i=1}^{s-1}\sum_{j=s}^{n-1}
\left[  
\left(\frac{1}{ k_{(i+1,n),+}}-\frac{1}{{ k_{(j+1,n),+}}}\right)
\frac{k_{s+}}{k_{(i+1,j),+}}
\sbraket{\eta k_{1,i}}\sbraket{k_{j+1,n}\eta}
\right.\nonumber\\
&\left.\times\bar\Sigma^+(-k_{1,i},k_1,\dots k_i)
\mathcal{X}^s(-k_{i+1,j},k_{i+1},\dots k_j)
\Sigma^+(-k_{j+1,n},k_{j+1},\dots k_n)
\right]\nonumber\\
&=\frac{ m(\sqrt 2 g)^{n-2}}{
\braket{12}\dots\braket{(n-1)n}}
\frac{\braket{\eta s}^2 \braket{1s}\braket{sn}}{\braket{\eta 1}\braket{\eta n}}
\,.
\label{eq:vertex+-+}
\end{align}
Details are again provided in appendix~\ref{app:details}.
In agreement with the SUSY-WI~\eqref{eq:neg-swi1} the 
 vertex~\eqref{eq:vertex+-+} vanishes for the choice
$\ket{\eta+}=\ket{s+}$.
The  new four-quark vertices with a helicity flip on one of the quark lines
indicated in~\eqref{eq:trafo+-+} are of the form:
 \begin{align}
L^{(n)}_{\bar\psi^+ B\dots \psi
\bar\psi B\dots \psi^+}=
\sum_{\sigma=\pm}&\sum_{s=2}^{n-1}
\int_{1\dots n} \left[
\mathcal{V}_{\bar\psi^{+}_1 B_2\dots \psi^{-\sigma}_s
\bar\psi_{s+1}^\sigma B_{s+2}\dots \psi_n^{+}} 
(\bar\psi^{+}_{k_1} B_{k_2}\dots \psi^{-\sigma}_{k_s})(
\bar\psi_{k_{s+1}}^\sigma B_{k_{s+2}}\dots \psi_{k_n}^{+})\right.\nonumber\\
&\left.+\frac{1}{N}
\mathcal{V}_{\bar\psi^{+}_1 B_2\dots \psi^{+}_s
\bar\psi_s^\sigma B\dots \psi_n^{-\sigma}} 
(\bar\psi^{+}_{k_1} B_{k_2}\dots \psi^{+}_{k_s})(
\bar\psi_{k_{s+1}}^\sigma B_{k_{s+2}}\dots \psi_{k_n}^{-\sigma})\right]\,.
 \end{align}
The calculation of the coefficients is identical to the previous case 
of the two-quark helicity flip vertex~\eqref{eq:vertex+-+}, 
up to the replacement
of $k_{s+}\mathcal{X}^s$ by $\mathcal{K}_{\pm}^s$ and a change of the
lower bound of the $j$-summation to $s+1$. The resulting vertices are
\begin{align}
  \mathcal{V}_{\bar \psi^+_1,B_{2},\dots , \psi_s^-,
\bar \psi_{s+1}^{+},\dots B_{n-1},\psi^+_n}
&=-\frac{ m(\sqrt 2 g)^{n-2}\braket{1s}\braket{(s+1)n}}{
\braket{12}\dots\braket{(n-1)n}}
\frac{\braket{\eta s}^2}{
\braket{\eta 1}\braket{\eta n}}\,, \\
  \mathcal{V}_{\bar \psi^+_1,B_{2},\dots , \psi_s^+,
    \bar \psi_{s+1}^{-},\dots B_{n-1},\psi^+_n}
&=\frac{ m(\sqrt 2 g)^{n-2}\braket{1s}\braket{(s+1)n}}{
\braket{12}\dots\braket{(n-1)n}}
\frac{\braket{\eta (s+1)}^2}{\braket{\eta 1}\braket{\eta n}}\,.
\end{align}
This completes  the derivation of the new vertices required in the CSW rules
for massive quarks.

\subsection{Diagrammatic rules and discussion}
\label{sec:rules}
For convenience, we collect the vertices derived in the previous subsection
and give some comments on their application and the structure
of the resulting diagrams.
The vertices needed for the computations of color-ordered partial amplitudes
in the conventions summarized in appendix~\ref{app:color} are obtained from the
vertex functions $\mathcal{V}$ derived in the previous subsection simply by
dropping the coupling-constant factors (and multiplying by $\ii$).

There is one subtlety for the CSW rules with internal fermion
lines~\cite{Bern:2004ba}. In the conventions used up to now all momenta were
treated as outgoing.  These are also the conventions used for the MHV
amplitudes in appendix~\ref{app:color}.  However, after summing over helicity
combinations of internal quark lines the CSW diagrams should reconstruct the
usual Dirac propagators where the momentum flows along the fermion line.
Therefore in the rules
given in the following the momenta of all outgoing anti-quarks (denoted by
$\psi$) will be reversed so they are treated as
 incoming quarks instead.  In the
conventions used in~\cite{Schwinn:2005pi} massless and massive spinors with
opposite momenta are related by $\ket{(-k)\pm}=\ii \ket{k\pm}$ and $\ii
u(k,\pm)= v(-k,\pm)$ if $k^0>0$.  To obtain vertices for incoming quarks one
therefore has to multiply the amplitude for outgoing anti-quarks by a factor
of $-\ii$ for every $\psi$ in addition to reversing the momenta (assuming that
the incoming quarks have positive energy).  It turns out that all vertices are
homogenous of degree $1$ in the spinors associated to quarks or anti-quarks
with a minus label and homogenous of degree $(-1)$ in the spinors associated
to quarks or anti-quarks with a plus label.  This implies that flipping the
anti-quark momenta gives a sign-change for each $\psi^+$ while there is no
change for $\psi^-$.

Taking the above conventions into account, we can now list all the
vertices for the massive CSW rules. 
The vertices present both for massless
 and massive quarks are given by off-shell continuations of the MHV amplitudes for two or four quarks in the fundamental representation:
\begin{align}
\label{eq:csw-glue}
  V_{\text{CSW}}(\bar B_1,B_2,\dots \bar B_{i},\dots, B_n)&=
\ii 2^{n/2-1}  \frac{ \braket{1i}^4}{
  \braket{12}\dots\braket{(n-1)n}\braket{n1}}\, ,\\
\label{eq:csw-2psi}
  V_{\text{CSW}}(\bar\psi_1^-,B_2,\dots \bar B_{i},\dots \psi_n^+)&=
\ii 2^{n/2-1}  \frac{ \braket{1i}^3\braket{in}}{
  \braket{12}\dots\braket{(n-1)n}\braket{n1}}\, , \\
\label{eq:csw-4psi1}
  V_{\text{CSW}}( \bar\psi_1^-,B_2,\dots \psi^+_{i},\bar\psi^-_{i+1}\dots \psi^+_n)&=-\ii 2^{n/2-1}  \frac{ \braket{1i}\braket{(i+1)n}\braket{1(i+1)}^2}{
  \braket{12}\dots\braket{(n-1)n}\braket{n1}}\, ,\\
\label{eq:csw-4psi2}
  V_{\text{CSW}}( \bar\psi_1^-,B_2,\dots \psi^-_{i},\bar\psi^+_{i+1}\dots \psi^+_n)&=-\ii 2^{n/2-1}  \frac{ \braket{1i}^3\braket{(i+1)n}}{
  \braket{12}\dots\braket{(n-1)n}\braket{n1}}\,.
\end{align}
The new vertices only present for massive quarks
are given by `helicity conserving' 
vertices proportional to $m^2$ generated from
the transformation of the mass term:
\begin{equation}
\label{eq:csw-m2}
\begin{aligned}
V_{\text{CSW}}(\bar \psi^-_1,B_{2},\dots B_{n-1},\psi^+_n)
&=\ii  2^{n/2-1} m^2 \frac{\braket{\eta 1} \braket{1n}}{
\braket{12}\dots\braket{(n-1)n }\braket{n\eta}}\; ,\\
V_{\text{CSW}}(\bar \psi^+_1,B_{2},\dots B_{n-1},\psi^-_n)
&=\ii  2^{n/2-1} m^2 \frac{\braket{1n}\braket{n\eta } }{\braket{\eta 1}
\braket{12}\dots\braket{(n-1)n }}\; ,
\end{aligned}
\end{equation}
helicity flip vertices for a single quark pair:
\begin{align}
V_{\text{CSW}}(\bar \psi^-_1,B_{2},\dots B_{n-1},\psi^-_n)
&=-\ii  2^{n/2-1}m\,
\frac{ \braket{1n}^2}{
\braket{12}\dots\braket{(n-1)n }}\, ,
\label{eq:csw-m-}\\ 
V_{\text{CSW}}(\bar \psi^+_1,B_{2},\dots ,\bar B_i,\dots \psi^+_n)&=
-\ii  2^{n/2-1}m\, \frac{  \braket{1i}\braket{in}}{
\braket{12}\dots\braket{(n-1)n}}
\frac{\braket{\eta i}^2}{\braket{\eta 1}\braket{\eta n}}\,,
\label{eq:CSW+-+}
\end{align}
and four-quark vertices that flip the  helicity of a single quark line:
\begin{equation}
\label{eq:4qflip}
\begin{aligned}
  V_{\text{CSW}}( \bar\psi_1^+,B_2,\dots \psi^-_{i},\bar\psi^+_{i+1}\dots \psi^+_n)&=\ii  2^{n/2-1}m\, \frac{  \braket{1i}\braket{(i+1)n}}{
\braket{12}\dots\braket{(n-1)n}}
\frac{\braket{\eta i}^2}{\braket{\eta 1}\braket{\eta n}}\, ,\\
  V_{\text{CSW}}( \bar\psi_1^+,B_2,\dots \psi^+_{i},\bar\psi^-_{i+1}\dots \psi^+_n)&=\ii  2^{n/2-1}m\, \frac{  \braket{1i}\braket{(i+1)n}}{
\braket{12}\dots\braket{(n-1)n}}
\frac{\braket{\eta (i+1)}^2}{\braket{\eta 1}\braket{\eta n}}\, .
\end{aligned}
\end{equation}
The propagators are given by $\ii/(p^2-m^2)$ for the quarks and
$\ii/p^2$ for the gluons and connect plus and minus labels of the vertices.
The color structure of the new vertices for massive quarks is identical to
those for massless quarks. Therefore the leading and sub-leading color
structures arise in the same way  as discussed at the end of subsection~\ref{sec:trafos}.
Note that the vertices presented in the revised version of~\cite{Ettle:2008ey}
involve non-canonically normalized fields so one has to take the normalization factors in~\eqref{eq:physical-amplitudes} into account to compare to their results.

For off-shell particles and external on-shell massive quarks the spinors
in these vertices are continued off-shell according to the prescription~\eqref{eq:continue} which amounts to the replacement
\begin{equation}
\label{eq:continue-spinor}
  \braket{1i}\rightarrow \frac{\braket{\eta+|\fmslash k_1|i+}}{\sbraket{\eta 1}}\end{equation}
for a massive or off-shell particle with momentum $k_1$ and a spinor
$\ket{i+}$ associated to a light-like momentum.
Since the prescription~\eqref{eq:continue-spinor} also has to be
applied for external massive quarks, the anti-holomorphic spinor
products like $\sbraket{\eta 1}$ introduced by the off-shell
continuation~\eqref{eq:continue-spinor} of external massive momenta
are not guaranteed to cancel out in the final amplitude.  This is in
contrast to the CSW rules for massless particles, where only spinor
products for internal off-shell momenta have to be continued and the
denominators of~\eqref{eq:continue-spinor} always cancel between different
vertices.  The remaining dependence of the massive quark amplitudes
on  $\ket{\eta-}$   (as well as
the explicit dependence of some vertices on $\ket{\eta+}$) reflects the
dependence on the quantization axis of the massive quark spin.
  Also note that---in contrast to the massless
case~\cite{Kosower:2004yz}---the off-shell continuation
using~\eqref{eq:continue} is not equivalent to the original
prescription~\cite{Cachazo:2004kj} that amounts to dropping the denominator
in~\eqref{eq:continue-spinor}.  For massive scalars~\cite{Boels:2007pj,Boels:2008ef},  $\eta$-independence of scattering
amplitudes follows from the independence of amplitudes on the vector
defining the gauge choice $\eta\cdot A=0$. In this case, both
prescriptions for the off-shell continuation are equivalent since
 all vertices are homogenous of degree zero in
spinors associated to massive scalars (see e.g.~\eqref{eq:csw-mass})
so the denominators from~\eqref{eq:continue-spinor} cancel as they must.

The structure of the amplitudes constructed using the above rules has
some similarities to the rules of~\cite{Schwinn:2005pi} reviewed in
section~\ref{sec:lc-vertices}. Defining 
the degree of a vertex or of an amplitude 
as the number of `$-$'-labels minus one, 
 only  vertices of degree zero and one 
occur. Furthermore, the degree of an amplitude is  the sum of the degrees of
the vertices.
Since all the vertices of the massless
CSW rules have degree one, exactly $d$ of these vertices contribute to
an amplitude with with $d+1$ `$-$'
labels. These massless MHV
vertices have to be dressed in all possible ways by the degree-zero $m^2$
 vertex~\eqref{eq:csw-m2}.  
In contrast, 
the number of helicity flip vertices~\eqref{eq:csw-m-}--\eqref{eq:4qflip} in a diagram is bounded in terms of the degree
of the amplitude~\cite{Schwinn:2005pi}.  For an amplitude with one
massive quark pair and the helicity configuration $\psi^+
\bar{\psi}^-$ the total number of helicity flips $f$ is bounded by the
degree $d$ of the amplitude as
\begin{equation}
 f_{+-}  \le  2 d  
\end{equation}
which follows from the fact that the helicity flips must occur in pairs and the
vertex~\eqref{eq:csw-m-} has degree one.  
Similarly for  the helicity configuration $\psi^+ \bar{\psi}^+$ ($\psi^- \bar{\psi}^-$)
the total number of flips is bounded by 
\begin{equation}
\label{eq:n-flip+}
 f_{\pm\pm}  \le  2 d \pm 1.    
\end{equation}
Therefore the
structure of diagrams in the massive CSW rules is intermediate to that in
the light-cone formalism of~\cite{Schwinn:2005pi} and the massless CSW rules: all gluonic
degree-zero vertices are eliminated by the field redefinition so that
the number of diagrams is smaller than in the light-cone formalism,
but in contrast to the massless CSW formalism the total number of
vertices is not fixed by the degree of the amplitude.  
\section{Simple applications and extension to different quark flavors}
\label{sec:applications}

In this section
we give some simple examples for the applications of the rules derived
in section~\ref{sec:derive} and present
the extension to amplitudes with quarks of different masses.
Subsection~\ref{sec:examples} contains some explicit examples for
the application of the rules to diagrams with up to five external legs, 
in subsection~\ref{sec:all-n} the structure of some all-multiplicity
amplitudes is discussed and it is shown that they satisfy the appropriate 
SUSY-WIs. In subsection~\ref{sec:flavors} the rules are generalized
to several quark flavors with different masses.

\subsection{Examples}
\label{sec:examples}

In order to check that the rules derived in the previous section 
give the correct scattering amplitudes for massive quarks, 
we will consider some simple explicit examples.

Amplitudes with a pair of massive quarks and
only positive helicity gluons are directly proportional
to scalar amplitudes due to the SUSY-WIs~\eqref{eq:conserve+++}--\eqref{eq:flip+++}.
In particular, the fact that the scalar three-point $m^2$ CSW vertex
$V_{\text{CSW}}(\bar \phi, B,\phi)$  obtained from~\eqref{eq:csw-mass} agrees with the corresponding vertex in the usual spinor-helicity
formalism if all external lines
 are on-shell~\cite{Boels:2007pj}
 (which is possible for complex external momenta) implies the
same property for the corresponding quark vertices, i.e. the
three-point vertices obtained from~\eqref{eq:csw-m2} and~\eqref{eq:csw-m-}.
Therefore no  `equivalence theorem violating' contributions~\cite{Ettle:2007qc} arise for these vertices,
 in agreement with the general discussion in~\cite{Boels:2008ef}.
Amplitudes with one negative helicity gluon are only determined by scalar
amplitudes for a particular choice of the reference spinors so they provide
more interesting tests of the rules. 
Both the helicity flip three point vertex obtained from~\eqref{eq:CSW+-+}
and  the MHV vertex~\eqref{eq:csw-2psi}
are in  agreement with the corresponding light-cone vertices~\eqref{eq:lcg-vertices} so no equivalence theorem evasion arises in this case either. 

\begin{figure}[t]
\centering
  \includegraphics[width=0.7\textwidth]{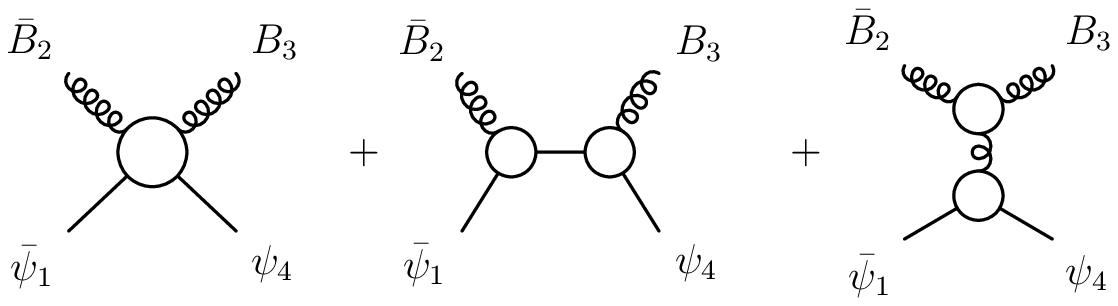}
  \caption{Topologies contributing to four-point amplitudes with a massive quark pair}
\label{fig:four-point}
\end{figure}

To demonstrate the application of the massive CSW rules,
 four point amplitudes with one
negative helicity gluon will be discussed in detail.
 In this case of course the amplitudes
can be simply obtained from Feynman diagrams; the results are
summarized in appendix~\ref{app:amplitudes}. 
  The three topologies contributing to the
massive CSW diagrams for general helicity combinations are shown in
figure~\ref{fig:four-point}.  The helicity flip
amplitude with two negative helicity quarks has degree two, so
according to the discussion in section~\ref{sec:rules} every diagram
must contain one flip-vertex~\eqref{eq:csw-m-} and one massless MHV
vertex.  Since there is no four-point vertex of degree two, only the
second and third diagram in~\ref{fig:four-point} contribute.  
Using the off-shell continuation~\eqref{eq:continue-spinor}
and applying momentum conservation one obtains
\begin{align}
  A_n( \bar \psi_1^-, \bar B_2, B_3, \psi^-_{4})
&=\frac{- 2 \braket{12}^2}{\braket{k_{1,2}1}}
\frac{\ii }{k_{1,2}^2-m^2} 
\frac{ m\braket{k_{1,2}4}^2}{\braket{k_{1,2}3}\braket{34}}
+\frac{-2  m\braket{14}^2}{\braket{1k_{2,3}}\braket{k_{2,3}4}}
\frac{\ii}{k_{2,3}^2}
\frac{\braket{k_{2,3}2}^3}{\braket{23}\braket{3k_{2,3}}}
\nonumber\\
&=2\ii m \left[
\frac{\braket{\eta+|\fmslash k_1|2+}
\sbraket{\eta 3}^2}{
2(k_1\cdot k_2)\sbraket{\eta 1}\sbraket{4\eta}\sbraket{2\eta}}
+\frac{\sbraket{\eta 3}^3}{
\sbraket{\eta 1}\sbraket{4\eta}
 \sbraket{\eta 2}\sbraket{32}}
\right]\nonumber\\
&=\frac{(2\ii m) \sbraket{3\eta}^2
\braket{2-|\fmslash k_1|3-}}{\sbraket{\eta 1}\sbraket{4\eta}
 \sbraket{32} 2(k_1\cdot k_2) }
\end{align}
where in the last step Dirac algebra was  used to combine the two terms.
The same expression can be obtained from the Feynman diagram result~\eqref{eq:feyn4}. 
As anticipated below~\eqref{eq:continue-spinor}, the off-shell continuation
has introduced a dependence on the anti-holomorphic spinor products
$\sbraket{\eta 1}$ and $\sbraket{\eta 4}$.

For the helicity flip amplitude for two positive helicity quarks there
is a contribution from the four-point vertex in figure~\ref{fig:four-point}
but in this case the diagram with the gluonic MHV vertex does not
contribute since it would require a quark vertex with only plus labels. 
The diagrams from the CSW rules simplify to
  \begin{align}
      A_n( \bar \psi_1^+, \bar B_2, B_3, \psi^+_{4})
&=\frac{-2 \ii m\braket{\eta 2}^2\braket{24}}{
\braket{\eta 1}\braket{23}\braket{34}\braket{\eta 4}}
+\frac{-\sqrt 2 \ii m\braket{\eta 2}^2}{\braket{\eta 1}\braket{\eta k_{1,2}}}
\frac{\ii}{k_{1,2}^2-m^2}
\frac{-\sqrt 2 \ii m^2\braket{\eta k_{1,2}}\braket{k_{1,2}4}}{
\braket{k_{1,2}3}\braket{34}\braket{\eta 4} }\nonumber\\
&=\frac{-2 \ii m\braket{\eta 2}^2}{
\braket{\eta 1}\braket{3+|\fmslash k_4|\eta+}\braket{\eta 4}}\left[
\frac{\braket{2-|\fmslash k_4|\eta+}}{\braket{23}}
-\frac{m^2\sbraket{\eta 3}}{2(k_1\cdot k_2)} \right]\nonumber\\
&=\frac{2 \ii m\braket{\eta 2}^2\braket{3+|\fmslash k_1|2+}}{
\braket{\eta 1}\braket{\eta 4}2(k_1\cdot k_2)\braket{23}}\,.
  \end{align}
In the last step we have used momentum conservation and Dirac algebra
to write $2(k_1\cdot k_2)\braket{2-|\fmslash k_4|\eta+}
=-\braket{2-|\fmslash k_1|3-}\braket{3+|\fmslash k_4|\eta+}
-m^2\braket{2-|\fmslash k_3|\eta-}$. This result can be seen to be identical
to that obtained from the Feynman result~\eqref{eq:feyn4}.

Finally, all topologies in figure~\ref{fig:four-point} contribute
to the helicity conserving amplitudes.
One finds, for example
\begin{equation}
\begin{aligned}
\label{eq:amp-+}
  A_4( \bar \psi_1^-, \bar B_2, B_3, \psi_{4}^+)
=&2\ii \frac{\braket{12}^2\braket{24}}{\braket{23}\braket{34}\braket{41}}
+\frac{-\sqrt 2 \ii m^2\braket{14}\braket{\eta 1}}{
\braket{1k_{2,3}}\braket{k_{2,3}4}\braket{\eta 4}}
\frac{\ii}{k_{2,3}^2}\frac{\sqrt 2 \ii\braket{ k_{2,3}2}^3}{\braket{23}\braket{3k_{2,3}}}\\
&+\frac{\sqrt 2\ii \braket{12}^2}{\braket{k_{1,2}1}}
\frac{\ii }{k_{1,2}^2-m^2} 
\frac{-\sqrt 2 \ii m^2\braket{k_{1,2}4}\braket{\eta k_{1,2}}}{
\braket{k_{1,2}3}\braket{34}\braket{\eta 4}}\,.
\end{aligned}
\end{equation}
For the choice $\ket{\eta-}=\ket{3-}$  the last two terms drop
out
and one finds in agreement with~\eqref{eq:feyn4}:
\begin{equation}
  A_4( \bar \psi_1^-, \bar B_2, B_3, \psi_{4}^+)|_{\ket{\eta-}=\ket{3-}}
=2\ii \frac{\braket{3+|\fmslash k_1|2+}^2}{2(k_3\cdot k_4)
 \braket{23}\sbraket{32}} \frac{\braket{\eta 1}
\braket{\eta-|\fmslash k_4|3-}}{\braket{\eta 4}\braket{\eta-|\fmslash k_1|3-}}
\end{equation}

\begin{figure}[t]
  \centering
 \includegraphics[width=0.9\textwidth]{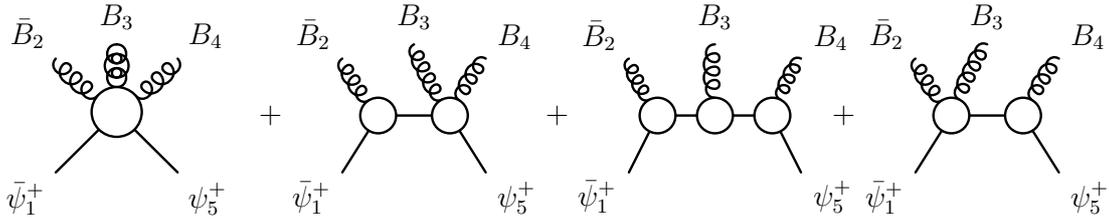}
  \caption{Topologies contributing to the
five-point helicity flip amplitude with a massive quark pair with  positive helicity and a negative helicity gluon adjacent to a massive quark.}
\label{fig:flip+5}
\end{figure}
As an example for a five point function consider the helicity flip amplitude
with two positive helicity quarks and a negative helicity gluon
adjacent to a quark. Since this is a degree zero amplitude, according
to~\eqref{eq:n-flip+} there can be only one helicity flip
vertex~\eqref{eq:CSW+-+} in each diagram, combined in all possible
ways with the vertex~\eqref{eq:csw-m2}.  As shown 
 in figure~\ref{fig:flip+5} there are four  contributing diagrams, 
in contrast to six diagrams in the usual color-ordered Feynman rules.
As discussed in~\cite{Boels:2007pj} the three-point vertex
with gluon $B_4$  in the last two 
diagrams
vanishes for the choice $\ket{\eta-}=\ket{4-}$. 
The remaining two diagrams give
\begin{equation}
   A_5(\bar Q_1^{+},g_2^-,g_{3}^+,g_{4}^+,Q_5^{+})=
\frac{2^{3/2}\ii m \braket{\eta 2}^2}{\braket{\eta 1}\braket{\eta 5}
\braket{34}y_{1,3}}
\left[ \frac{\braket{2-|\fmslash k_5|4-}}{\braket{23}}
+\frac{m^2\braket{4+|\fmslash k_{1,2}|3+}\sbraket{34}}{y_{1,2}
\braket{4+|\fmslash k_5|3+}}
\right]\,.
\end{equation}
with $y_{i,j}=k_{i,j}^2-m^2$.
This agrees numerically with the result~\eqref{eq:npt-++} obtained from on-shell recursion relations.

\subsection{Structure of simple all-multiplicity amplitudes and SUSY-WIs}
\label{sec:all-n}
In this subsection  a recursive construction  of amplitudes for a massive quark pair and an arbitrary number of positive helicity gluons and no or one negative helicity gluon is used to show that 
these amplitudes manifestly satisfy the SUSY-WIs reviewed in appendix~\ref{sec:swi} if they are calculated using the massive CSW rules.

\begin{figure}[t]
  \centering
 \includegraphics[width=0.65\textwidth]{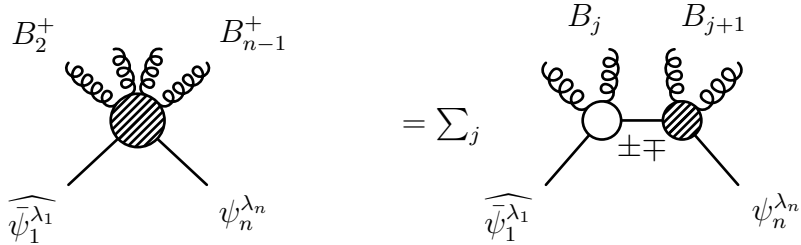}
  \caption{Recursive construction of amplitudes with only positive helicity gluons. The grey blobs denote amplitudes with one off-shell leg and the white blob denotes the vertex~\eqref{eq:csw-m2} or~\eqref{eq:csw-m-}}
\label{fig:n+}
\end{figure}

Amplitudes with only positive-helicity gluons
 can be obtained recursively from a relation involving 
currents with one off-shell quark (denoted by a hat), as shown in figure~\ref{fig:n+}:
\begin{multline}
\label{eq:csw-recursion}
   A_n( \widehat{\bar \psi^{\sigma_1}_1},B_2,... ,\psi^{\sigma_n}_{n})\\
= \sum\limits_{j=2}^{n-1} \sum_{\sigma=\pm}
V_{j+1,\text{CSW}}(\bar \psi_1^{\sigma_1},B_2...,B_{j}, 
\psi^{\sigma}_{k_{1,j}})\,
 \frac{\ii}{k_{1,j}^2-m^2}
 A_{n-j+1}(\widehat{\bar \psi^{-\sigma}_{k_{1,j}}},B_{j+1},\dots 
\psi^{\sigma_n}_{n}) \,.
\end{multline}
Recall that gluons and $\bar\psi$ fields are treated as outgoing and $\psi$
fields as incoming.  The two-point function is defined as $ A_2(\widehat{\bar
  \psi_{p}^{\pm}},\psi_p^{\mp}) =(-\ii) (p^2-m^2)$.
Using~\eqref{eq:csw-recursion} iteratively, the $n$-particle amplitude is
expressed as a sum of diagrams with $1,2,\dots n-2$ massive quark-vertices,
~\eqref{eq:csw-m2} or~\eqref{eq:csw-m-} depending on the helicity
configuration, summed over all possible distributions of the gluons.  Since
there is no vertex with only $B$-fields and quarks with plus labels, the
all-plus amplitude vanishes.
 The scalar amplitudes constructed using the vertex~\eqref{eq:csw-mass} and a recursive definition analogous to~\eqref{eq:csw-recursion} have been shown to satisfy the appropriate on-shell recursion relation~\cite{Boels:2007pj} so this
property is inherited by the quark amplitudes, once it is demonstrated that
these satisfy the appropriate SUSY-WIs~\eqref{eq:conserve+++} and~\eqref{eq:flip+++}. Since all massive CSW vertices contain only holomorphic spinor products, the twistor-space properties of massive quark amplitudes are analogous to those
of scalars~\cite{Boels:2007pj}, i.e. each term in the sum in~\eqref{eq:csw-recursion} localizes on a set of lines in twistor space, connected by scalar propagators.

For the helicity conserving amplitudes $A_n(\bar \psi^{+}_1,B_2,...
,\psi^{-}_{n})$ only the $\sigma=-$ term in the sum over helicities
contributes so only the vertices~\eqref{eq:csw-m2} appear.  Since these are
related to the scalar vertices by the relation~\eqref{eq:csw+++} it follows by
an inductive argument as in~\cite{Schwinn:2006ca} that the quark
scattering amplitudes are related by the SUSY-WI~\eqref{eq:conserve+++} to the
corresponding scalar amplitudes. 
 For the helicity flip amplitudes
$A_n(\bar \psi^{-}_1,B_2,... ,\psi^{-}_{n})$ both terms
in the sum over helicities contribute.
 Assuming the one-particle off-shell currents
with up to $n-1$ legs satisfy the SUSY-WI~\eqref{eq:flip+++}
 and using the relation of the
vertices~\eqref{eq:csw-m2} and~\eqref{eq:csw-m-} to the scalar vertex~\eqref{eq:csw-mass} one obtains from~\eqref{eq:csw-recursion}
\begin{equation}
\begin{aligned}
   A_n( \widehat{\bar \psi^{-}_1},B_2,... ,\psi^{-}_{n})=&
 \sum\limits_{j=2}^{n-1} 
V_{j+1,\text{CSW}}(\bar \phi_1,B_2...,B_{j}, 
\phi_{-k_{1,j}}) \frac{i}{k_{1,j}^2-m^2} \times\\
&\left(\frac{\braket{\eta 1}}{\braket{\eta k_{1,j}}}
\frac{\braket{k_{1,i} n}}{m}
+\frac{\braket{1k_{1,i}}}{m}\frac{\braket{\eta n}}{\braket{\eta k_{1,j}}}
\right)
 A_{n-j+1}(\widehat{\bar \phi_{k_{1,j}}},B_{j+1},\dots \phi_{n}) \\
&=\frac{\braket{1n}}{m}A_n( \widehat{\bar \phi_1},..., ,\phi_{n}))\,.
\end{aligned}
\end{equation}
Therefore the amplitude satisfies~\eqref{eq:flip+++} by induction,
 as was to be shown.

\begin{figure}[t]
  \centering
 \includegraphics[width=\textwidth]{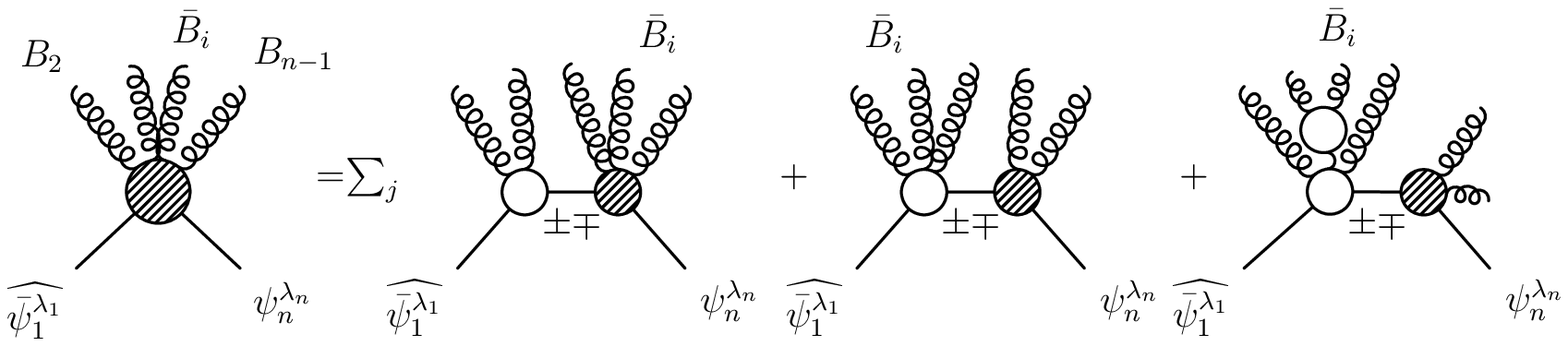}
  \caption{Recursive construction of amplitudes with one negative helicity gluon. Grey blobs denote amplitudes with one off-shell leg and white blobs denote CSW vertices.}
\label{fig:n+-}
\end{figure}

The structure of amplitudes with one negative-helicity gluon is more involved. 
As shown in figure~\ref{fig:n+-}
 there are three different types of contributions to the
off-shell recursion relations:
\begin{multline}
\label{eq:csw-recursion-2}
   A_n( \widehat{\bar \psi^{\sigma_1}_1},B_2,...,\bar B_i ,
B_{i+1},\dots\psi^{\sigma_n}_{n})\\
=\sum_{\sigma=\pm}\Biggl( \sum\limits_{j=2}^{i-1} 
V_{j+1,\text{CSW}}(\bar \psi_1^{\sigma_1},B_2...,
 \psi^{\sigma}_{k_{1,j}})\,
 \frac{\ii}{k_{1,j}^2-m^2} 
 A_{n-j+1}(\widehat{\bar \psi^{-\sigma}_{k_{1,j}}},B_{j+1},\dots ,
\bar B_i,\dots,\psi^{\sigma_n}_{n}) \\
+ \sum\limits_{j=i}^{n-1} 
V_{j+1,\text{CSW}}(\bar \psi_1^{\sigma_1},B_2...,\bar B_i,
\dots,B_{j}, \psi^{\sigma}_{k_{1,j}})\,
 \frac{\ii}{k_{1,j}^2-m^2} 
 A_{n-j+1}(\widehat{\bar \psi^{-\sigma}_{k_{1,j}}},B_{j+1},\dots ,
\psi^{\sigma_n}_{n})\\
+\sum_{j=2}^{i}\sum_{k=i}^{n-1}\sum_{l=k+1}^{n-1}
V_{l+j-k+1,\text{CSW}}(\bar\psi_1^{\sigma_1},B_2...,B_{j-1}
 B_{k_{j,k}},B_{k+1},\dots,B_l,\psi^{\sigma}_{k_{1,l}})\times\\
 \frac{\ii}{k_{j,k}^2}
V_{k-j+2\text{CSW}}(\bar B_{-k_{j,k}} B_j,\dots \bar B_i,\dots B_{k}) 
\frac{\ii}{k_{1,l}^2-m^2}
 A_{n-l+1}(\widehat{\bar \psi^{-\sigma}_{k_{1,l}}},B_{l+1},\dots,\psi^{\sigma_n}_{n}) 
\Biggr)\,.
\end{multline}
The case $j=k=i$ in the above sum has to be dropped since it would lead to 
a two point function with two negative helicity gluons.
The amplitude with one negative helicity gluon enters itself in the
 first term on the right-hand side that contributes only for $i\geq 3$.
Note that in contrast to 
 the conventional Berends-Giele relations~\cite{Berends:1987me} 
 only MHV vertices and not the much more complicated off-shell MHV currents are
required in~\eqref{eq:csw-recursion-2}\footnote{I thank German Rodrigo for
  stressing this point.}. 

The expression~\eqref{eq:csw-recursion-2} for the amplitudes
with one negative helicity gluon can be used to check the
SUSY-WIs~\eqref{eq:neg-swi1} and~\eqref{eq:neg-swi2}.
For the helicity flip amplitude with two positive helicity quarks 
the sum over quark helicities collapses and the last term in~\eqref{eq:csw-recursion-2} does not contribute so the recursion simplifies to
\begin{multline}
\label{eq:csw-recursion-flip}
   A_n( \widehat{\bar \psi^{+}_1},B_2,...,\bar B_i ,
B_{i+1},\dots\psi^{+}_{n})\\
= \sum\limits_{j=2}^{i-1} 
V_{j+1,\text{CSW}}(\bar \psi_1^{+},B_2...,
 \psi^{-}_{k_{1,j}})\,
 \frac{\ii}{k_{1,j}^2-m^2} 
 A_{n-j+1}(\widehat{\bar \psi^{+}_{k_{1,j}}},B_{j+1},\dots ,
\bar B_i,\dots,\psi^{+}_{n}) \\
+ \sum\limits_{j=i}^{n-1} 
V_{j+1,\text{CSW}}(\bar \psi_1^{+},B_2...,\bar B_i,
\dots,B_{j}, \psi^{+}_{k_{1,j}})\,
 \frac{\ii}{k_{1,j}^2-m^2} 
 A_{n-j+1}(\widehat{\bar \psi^{-}_{k_{1,j}}},B_{j+1},\dots ,
\psi^{+}_{n})\,.
\end{multline}
After re-inserting the recursive expression for the amplitude in the first term on the right-hand side, it is seen that all terms involve the helicity flip vertex~\eqref{eq:CSW+-+}. Since this vertex vanishes for the choice of the
reference spinor  $\ket{\eta+}=\ket{i+}$, the amplitude automatically satisfies the SUSY-WI~\eqref{eq:neg-swi1}.

For the helicity-conserving amplitude entering the SUSY-WI~\eqref{eq:neg-swi2}
the sum over helicities also simplifies for the choice  $\ket{\eta+}=\ket{i+}$ 
and one obtains (here it is understood that the same reference spinor is used
everywhere on the right-hand side)
\begin{multline}
   A_n( \widehat{\bar \psi^{+}_1},B_2,...,\bar B_i ,
B_{i+1},\dots\psi^{-}_{n})|_{\ket{\eta+}=\ket{i+}}\\
= \sum\limits_{j=2}^{i-1} 
V_{j+1,\text{CSW}}(\bar \psi_1^{+},B_2...,
 \psi^{-}_{k_{1,j}})\,
 \frac{\ii}{k_{1,j}^2-m^2} 
 A_{n-j+1}(\widehat{\bar \psi^{+}_{k_{1,j}}},B_{j+1},\dots ,
\bar B_i,\dots,\psi^{-}_{n}) \\
+ \sum\limits_{j=i}^{n-1} 
V_{j+1,\text{CSW}}(\bar \psi_1^{+},B_2...,\bar B_i,
\dots,B_{j}, \psi^{-}_{k_{1,j}})\,
 \frac{\ii}{k_{1,j}^2-m^2} 
 A_{n-j+1}(\widehat{\bar \psi^{+}_{k_{1,j}}},B_{j+1},\dots ,
\psi^{-}_{n})\\
+\sum_{j=2}^{i}\sum_{k=i}^{n-1}\sum_{l=k+1}^{n-1}
V_{l+j-k+1,\text{CSW}}(\bar \psi_1^{+},B_2...,B_{j-1}
 B_{k_{j,k}},B_{k+1},\dots
 \psi^{-}_{k_{1,l}})\times\\
 \frac{\ii}{k_{j,k}^2}
V_{k-j+2\text{CSW}}(\bar B_{-k_{j,k}} B_j,\dots \bar B_i,\dots B_{k}) 
\frac{\ii}{k_{1,l}^2-m^2}
 A_{n-l+1}(\widehat{\bar \psi^{+}_{k_{1,l}}},B_{l+1},\dots,\psi^{-}_{n}) \,.
\end{multline}
In the second term, the helicity flip vertex~\eqref{eq:CSW+-+} has dropped out
due to the choice of reference spinor. 
The corresponding amplitude for scalars, 
$ A_n( \widehat{\bar \phi_1},B_2,...,\bar B_i ,
B_{i+1},\dots\phi_{n})$ is given by an identical expression with the obvious
replacement $\psi\to\phi$ everywhere.

It is easily seen by induction that the $n$-particle amplitude
satisfies the SUSY-WI~\eqref{eq:neg-swi2} if it is satisfied for the
amplitudes with up to $n-1$ external particles.  For instance, in the
first term on the right-hand side, the $m^2$ CSW
vertex~\eqref{eq:csw+++} is related to the scalar
vertex~\eqref{eq:csw-mass} by a factor $\braket{k_{1,j}i}/\braket{1i}$
while the lower point amplitudes by assumption satisfy the
identity~\eqref{eq:neg-swi2}. This combines to an over-all factor
$\braket{ni}/\braket{1i}$, as required.  In the second term on the
right-hand side the quark MHV-vertex~\eqref{eq:csw-2psi} is related to
the corresponding scalar vertex~\cite{Georgiou:2004wu} by a factor
$\braket{k_{1,i}i}/\braket{1i}$ while the validity of the
relation~\eqref{eq:conserve+++} for the off-shell amplitudes with
positive helicity gluons was just demonstrated below
eq.~\eqref{eq:csw-recursion} so again this term has the form required
by~\eqref{eq:neg-swi2}.  The third term works out analogously.

\subsection{Different quark flavors}
\label{sec:flavors}
As an extension of the rules for a single massive quark discussed up to now,
consider 
QCD with $N_f$ flavors of quarks with different masses:
\begin{equation}
\label{eq:qcd-flavor}
\mathcal{L}_{\text{QCD}}=  -\frac{1}{2}\tr[F^{\mu\nu}F_{\mu\nu}]
+\sum_{a=1}^{N_f}\bar\Psi_a(\ii\fmslash
D-m_a)\Psi_a\,.
\end{equation}
Of particular interest
for the purpose of phenomenology is the case of a single massive quark and 
the remaining quarks massless, that is relevant for the calculation 
of amplitudes for the production of a pair of top quarks and several jets.
The derivation of the light-cone Lagrangian goes through as 
in section~\ref{sec:light-cone}, the only difference being the equation of
motion of the $A_-$ component of the gluon~\eqref{eq:eom} that now includes
a sum over the quark flavors:
\begin{equation}
\label{eq:eom-flavor}
    A_{-,ij}=\frac{1}{\partial_+^2}
\left([D_\perp,\partial_+A_{\perp,ij}]
+\sum_a \frac{g}{2}\left( \bar\chi_{a,j} \fmslash n_+\chi_{a,i} 
-\frac{1}{N}\delta_{ij}(\bar\chi_a\fmslash n_+\chi_a) \right)\right)\,.
\end{equation}
Eliminating $A_-$ by the equation of motion then introduces off-diagonal 
four-quark terms in the Lagrangian~\eqref{eq:light-cone-lag}
 while all other terms are flavor-diagonal:
\begin{align}
\mathcal{L}=\mathcal{L}_{A_\perp}&+
\sum_a\bar\chi_a \left[\ii \partial_-+ 
(\ii \fmslash D_\perp-m_a)\frac{1}{2\ii\partial_+}(\ii \fmslash D_\perp+m_a)
+g \left(\frac{1}{\partial_+^2}[D_\perp,\partial_+A_{\perp}]\right)
\right]\fmslash n_+\chi_a \nonumber\\
&+
\sum_{a,b}\left(\frac{g}{2}\right)^2(\bar\chi_{a} \fmslash n_+\chi_{a})\otimes
\frac{1}{\partial_+^2}(\bar\chi_{b} \fmslash n_+\chi_{b})\,.
\label{eq:light-cone-lag-flavor}
\end{align}
The only modification required in the
derivation of the rules in section~\ref{sec:rules} 
is the expression for  the  
conjugate gluon momentum $A_{\bar z}$~\eqref{eq:bbar-em-quark} 
that now contains a sum over quark flavors:
\begin{equation}
  \label{eq:bbar-em-flavor}
\left.  p_+ (A_{p,\bar z})\right|_{\bar\psi\psi}
=\sum_{a=1}^{N_f}\sum_{n=2}^\infty
\sum_{s=1}^{n-1}\sum_{\lambda=\pm}
 \int_{1\dots n}
\mathcal{K}_\lambda^s(p,k_1,\dots k_n)
\left(B_{-k_{1}}\dots \psi^{-\lambda}_{a,-k_s}\right)\otimes
\left(\bar\psi^\lambda_{a,-k_{s+1}}\dots B_{-k_n}\right)\,.
\end{equation}
Since the field redefinitions used in the derivation of the rules
are independent on the mass, the coefficients $\mathcal{K}$ are the same 
for all quark flavors.

The resulting CSW rules will therefore consist of $N_f$ copies of the
2-quark vertices from section~\ref{sec:rules} with the obvious
replacement $m\to m_a$ for each flavor. The only modification of the
rules arises for the four-quark vertices that become off-diagonal in
the flavors. 
 For the massless MHV four-quark vertices, the new
structure arises both because of the modification of the four-quark
term in~\eqref{eq:light-cone-lag-flavor} and the $A_{\bar z}$
transformation~\eqref{eq:bbar-em-flavor}. The corresponding term in
the CSW Lagrangian~\eqref{eq:csw-lag-q4} is modified to
\begin{align}
L^{(n)}_{\bar\psi B\dots \psi
\bar\psi B\dots \psi}=
\sum_{a,b,\lambda,\sigma}&\sum_{j=2}^{n-1}
\int_{1\dots n}\!\!\left[
\mathcal{V}_{\bar\psi^{\lambda}_{a,1} B_2\dots \psi^{-\sigma}_{b,j}
\bar\psi_{b,j+1}^\sigma B_{j+2}\dots \psi_{a,n}^{-\lambda}} 
(\bar\psi^{\lambda}_{a,k_1} B_{k_2}\dots \psi^{-\sigma}_{b,k_j})(
\bar\psi_{b,k_{j+1}}^\sigma, \dots \psi_{a,k_n}^{-\lambda})\right.\nonumber\\
&\left.+\frac{1}{N}
\mathcal{V}_{\bar\psi^{\lambda}_{a,1} B_2\dots \psi^{-\lambda}_{a,j}
\bar\psi_{b,j}^\sigma B\dots \psi_{b,n}^{-\sigma}} 
(\bar\psi^{\lambda}_{a,k_1} B_{k_2}\dots \psi^{-\lambda}_{a,k_j})(
\bar\psi_{b,k_{j+1}}^\sigma, \dots \psi_{b,k_n}^{-\sigma})\right]\,.
\label{eq:csw-lag-q4-flavor}
\end{align}
The vertex functions are unchanged and given by the four-quark MHV amplitudes 
as before.
The four-point helicity flip vertices~\eqref{eq:4qflip} become off-diagonal in 
the flavors because of the 
sum over quark flavors in the expression of $A_{\bar z}$:
 \begin{align}
L^{(n)}_{\bar\psi^+ B\dots \psi
\bar\psi B\dots \psi^+}=
&\sum_{a,b,\sigma}\sum_{j=2}^{n-1}
\int_{1\dots n} \!\!\left[
\mathcal{V}_{\bar\psi^{+}_{a,1} B_2\dots \psi^{-\sigma}_{bj}
\bar\psi_{b,j+1}^\sigma B_{j+2}\dots \psi_{a,n}^{+}} 
(\bar\psi^{+}_{a,k_1} B_{k_2}\dots \psi^{-\sigma}_{b,k_j})(
\bar\psi_{b,k_{j+1}}^\sigma\dots \psi_{a,k_n}^{+})\right.\nonumber\\
&\left.+\frac{1}{N}
\mathcal{V}_{\bar\psi^{+}_{a,1} B_2\dots \psi^{+}_{a,j}
\bar\psi_{b,j}^\sigma B\dots \psi_{b,n}^{-\sigma}} 
(\bar\psi^{+}_{a,k_1} B_{k_2},\dots \psi^{+}_{a,k_j})(
\bar\psi_{b,k_{j+1}}^\sigma ,\dots \psi_{b,k_n}^{-\sigma})\right]\,.
 \end{align}
Again the vertex coefficients are unchanged up to the replacement $m\to m_a$:
\begin{equation}
\label{eq:CSW-4q-flip}
\begin{aligned}
  V_{\text{CSW}}( \bar\psi_{a,1}^+,B_2,\dots
  \psi^-_{b,i},\bar\psi^+_{b,i+1}\dots \psi^+_{a,n})&=
\ii  2^{n/2-1} m_a \frac{  \braket{1i}\braket{(i+1)n}}{
\braket{12}\dots\braket{(n-1)n}}
\frac{\braket{\eta i}^2}{\braket{\eta 1}\braket{\eta n}}\,,\\
  V_{\text{CSW}}( \bar\psi_{a,1}^+,B_2,\dots \psi^+_{b,i},\bar\psi^-_{b,i+1}\dots \psi^+_{a,n})&=\ii  2^{n/2-1} m_a\frac{  \braket{1i}\braket{(i+1)n}}{
\braket{12}\dots\braket{(n-1)n}}
\frac{\braket{\eta (i+1)}^2}{\braket{\eta 1}\braket{\eta n}}\,,
\end{aligned}
\end{equation}
with the same expressions for the color suppressed vertices. 

As a simple check of these rules,
 one can calculate four-quark 
amplitudes with one heavy quark pair $\bar QQ$ and one massless quark pair
$q\bar q$. 
It is easily seen that the only contribution to helicity flip 
amplitudes with two positive helicity
massive quarks are the four-point vertices
 obtained from~\eqref{eq:CSW-4q-flip} and that the results 
agree with the expression~\eqref{eq:4q-flip}
obtained from a Feynman diagram calculation. 

The helicity conserving amplitude receives contributions from the vertex~\eqref{eq:csw-2psi} and a diagram with two cubic vertices:
\begin{equation}
   A_4(\bar Q_1^+,q_2^+,\bar{q}_3{}^-,Q_4^-)
   =-\ii 2^{3/2}\left[ \frac{ \braket{34}^2}{\braket{23}\braket{41}}
+
\frac{m^2\braket{\eta 4}\braket{14}}{\braket{\eta 1}
\braket{1k_{2,3}}\braket{k_{2,3}4}} 
\frac{1}{k_{2,3}^2} 
\frac{\braket{3(-k_{2,3})}^2}{\braket{23}}\right]
\end{equation}
where in this case $k_{2,3}=k_2-k_3$. For the choice $\ket{\eta-}=\ket{2-}$ 
the second term vanishes and one gets
\begin{equation}
    A_4(\bar Q_1^+,q_2^+,\bar{q}_3{}^-,Q_4^-)
   =-\ii 2^{3/2} \frac{ \braket{3-|\fmslash k_4|2-}\braket{\eta-|\fmslash k_1|2-}}{\braket{\eta 1}\sbraket{42}\braket{23}\sbraket{23}}
\end{equation}
in agreement with the Feynman diagram result~\eqref{eq:4q-conserve}
 for this choice of $\ket{\eta-}$.

\section{Conclusions and outlook}
In this paper a method to derive CSW-like rules for colored massive particles
introduced previously for massive scalars~\cite{Boels:2007pj,Boels:2008ef} was
extended to massive quarks.
Simple rules were obtained by using the same auxiliary 
spinors to define
 the off-shell continuation of spinor products and the quantization axis of the heavy quark spin, avoiding complications encountered in an earlier
 proposal~\cite{Ettle:2008ey}.
The structure of the resulting tree diagrams was discussed and
 several simple examples for the application of the rules have been given. 
The rules have also been extended to amplitudes with quarks of different masses.

The rules presented here could be useful in several respects.
  As shown in~\cite{Boels:2008ef} the CSW rules for a massive scalar
  can in principle be applied to calculate the rational part of gluon
  amplitudes in pure Yang-Mills theory.  
 The contribution of a (massless) quark loop to the four-point amplitude with only positive helicity gluons can be obtained by a calculation identical to the one in~\cite{Boels:2008ef}.
It would  be
  interesting to see if the rules discussed in the present paper can
  also be useful in one-loop calculations of amplitudes for the production
  of top-quarks and jets at hadron colliders. 
The CSW rules for massive quarks could also be
  suitable for the simplification of proofs of on-shell recursion relations for 
  massive quarks that turn out to be rather tedious for
  some helicity combinations~\cite{Schwinn:2007ee}.
Finally, the fact that the new vertices for massive quarks are proportional 
to the mass suggests a possibility to derive a systematic expansion of scattering amplitudes in the quark mass.

The derivation of the rules was based on applying field redefinitions
obtained in the framework of canonical
transformations~\cite{Ettle:2006bw,Ettle:2008ey} and twistor
methods~\cite{Boels:2008ef,Boels:2008du} to the QCD Lagrangian in
light-cone gauge. It would be interesting to give a direct
construction of an action in twistor
space~\cite{Mason:2005zm,Boels:2007qn} that leads to the same rules as
the `twistor inspired' derivation given here.  This is not entirely
obvious since the lifting formulas for quarks to twistor
space~\cite{Boels:2008du} are somewhat more complicated than the one
for scalars~\cite{Boels:2008ef}.  It is also worthwhile to revisit the
CSW rules for electroweak currents~\cite{Bern:2004ba} in the light of
the derivation of the CSW rules using canonical field redefinitions.
Similarly to the related example of CSW rules for an effective
Higgs-gluon coupling~\cite{Boels:2008ef} a derivation of these rules
in the twistor Yang-Mills approach results in additional vertices not
presented in the previous literature~\cite{Boels:private}.  The CSW
rules for massive quarks are also expected to generalize to a direct
coupling of the Higgs boson to top quarks. This might be useful in
order to go beyond the large top mass limit used in previous
applications of the CSW rules to Higgs boson scattering amplitudes.

\section*{Acknowledgments}
I thank Rutger Boels and James Ettle for comments on the manuscript and German Rodrigo
for useful discussions.
This work was supported by the 
Bundesministerium f\"ur Bildung und Forschung (BMBF),  grant 05HT6PAA. 

\appendix
\section{Color decomposition and MHV amplitudes}
\label{app:color}

This appendix summarizes the conventions for the decomposition of the full
amplitude ${\cal A}_n$ into gauge invariant partial amplitudes
$A_n$\cite{Berends:1987me,Mangano:1990by} and the results for the MHV
amplitudes with one or two pairs of massless quarks.  In this appendix
the standard conventions are used where all momenta are treated as
outgoing.  For amplitudes with a single pair of quarks a suitable
decomposition is given by
\begin{multline}
\label{eq:color}
 {\cal A}_{n}(\bar{Q}_1,g_2,g_3,. ..,g_{n-1},Q_{n}) \\
=  
 g^{n-2} \sum_{\sigma\in S_{2,n-1}}  \; \left(
 T^{a_{\sigma(2)}} ... T^{a_{\sigma(n-1)}} \right)_{i_1i_n}
 A_{n}\left(\bar{Q}_1, g_{\sigma(2)}, ..., g_{\sigma(n-1)}, Q_n \right)\,.
\end{multline}
Here $S_{2,n-1}$ denotes the permutations of the $n-2$ elements
in the set  $(2,\dots n-1)$.
The partial amplitudes are calculated from color-ordered diagrams using
Feynman rules given e.g. in~\cite{Schwinn:2005pi}.
The MHV amplitudes with a pair of massless quarks, 
one negative  helicity gluon and an arbitrary number of positive helicity
gluons are given by
\begin{align}
  A_{n,\text{MHV}}(\bar Q_1^-,g_2^+,\dots g^-_{i},\dots Q_n^+)&=
-\ii 2^{n/2-1}  \frac{ \braket{1i}^3\braket{in}}{
  \braket{12}\dots\braket{(n-1)n}\braket{n1}}\,,\\
  A_{n,\text{MHV}}(\bar Q_1^+,g^+_2,\dots g^-_{i},\dots Q_n^-)&=
  \ii 2^{n/2-1}  \frac{ \braket{1i}\braket{in}^3}{
  \braket{12}\dots\braket{(n-1)n}\braket{n1}}\,.
\end{align}

For amplitudes with two quark pairs the color structure is more complicated
since contributions suppressed by the number of colors $N$ also have to be
taken into account.
For two different quark flavors $Q$ and $q$ the decomposition can be
written as 
\begin{multline}
\label{eq:color-4q}
 {\cal A}_{n+4}(\bar Q_{\bar p},Q_{p},\bar q_{\bar k},q_{k}, g_1,\dots g_n) 
= \frac{g^{n-2}}{2} \sum_{i=0}^{n} \sum_{\sigma\in S_{1,i}} 
\!\!\!\!\!\!\!\!\sum_{\;\;\;\;\;\;\;\sigma\in S_{i+1,n}}\\ 
\times\Bigl[ \left(
 T^{a_{\sigma(2)}} ... T^{a_{\sigma(i)}} \right)_{i_{\bar p}j_k}
\; \left(
 T^{a_{\sigma(i+1)}} ... T^{a_{\sigma(n)}} \right)_{j_{\bar k}i_p}
 A_{n}(\bar Q_{\bar p},g_1,,\dots g_i,q_{k},\bar q_{\bar k},g_{i+1}, ...,
g_{n},Q_{p}) \\
-\frac{1}{N}  
 \; \left(
 T^{a_{\sigma(2)}} ... T^{a_{\sigma(i)}} \right)_{i_{\bar p}i_p}
\; \left(
 T^{a_{\sigma(i+1)}} ... T^{a_{\sigma(n)}} \right)_{ j_{\bar k}j_k}
B_{n}(\bar Q_{\bar p},g_1,,\dots g_i, Q_{p};\bar q_{\bar k},g_{i+1}, ...,g_{n},q_{ k})\Bigr]
\,.
\end{multline}
In the cases $i=0$ and $i=n$ one of the strings of generators reduces to
a Kronecker delta. For amplitudes with two pairs of identical quark
flavors one has to subtract the right hand side after exchanging
$Q_p\leftrightarrow q_k$.  

The sub-leading color structures arise because of the color Fierz-identity
\begin{equation}
\label{eq:fierz}
  T^a_{ij}T^a_{kl}=\frac{1}{2}
\left(\delta_{il}\delta_{jk}-\frac{1}{N}\delta_{ij}\delta_{kl}\right)
\end{equation}
It is convenient to discuss this
 in the color-flow representation (see e.g. the lectures by Dixon in~\cite{Mangano:1990by})
 where gluons are depicted by double lines and
quarks by  single lines. The color Fierz-identity~\eqref{eq:fierz} implies that a gluon propagator connecting two quark pairs decomposes into  a leading color $U(N)$ piece and a color-suppressed $U(1)$ piece:
\begin{equation}
\label{eq:color-propagator}
\includegraphics[width=0.65\textwidth]{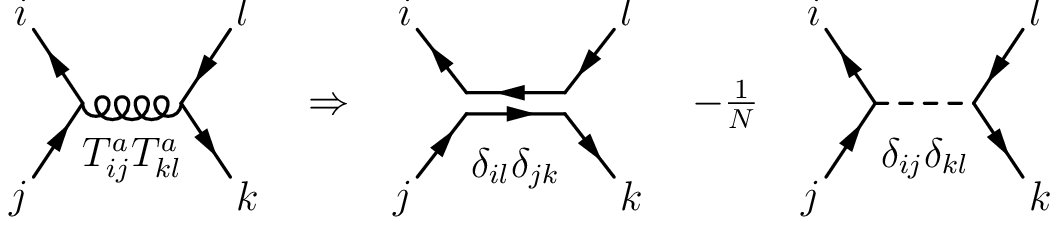}
\end{equation}

The color-ordered four-quark MHV amplitudes
for amplitudes with two different (massless) quark flavors contain
only positive helicity gluons and two negative helicity quarks. They
are given by
\begin{equation}
   A_{n}(\bar Q_{1},g^+_2,,\dots g^+_{i-1}, q_{i},\bar q_{i+1},
   g^+_{i+2}, ...,g^+_{n-1}, Q_{n})
   =-\ii 2^{n/2-1} 
 \frac{(-1)^{\sigma_{j_1j_2}}\braket{j_1j_2}^2\braket{1i }\braket{(i+1)n} }{
  \braket{12}\dots\braket{(n-1)n}\braket{n1}}
\end{equation}
where $j_{1}$ and $j_2$ are the two negative helicity quarks
and $\sigma_{1,i+1}=\sigma_{i,n}=0$, $\sigma_{1,i}=\sigma_{i+1,n}=1$.
The expression for the color suppressed MHV partial-amplitudes $B_n$
is identical (up to the obvious relabeling of the momenta).

\section{Supersymmetric Ward identities}
\label{sec:swi}
Scattering amplitudes with massive quarks can be related through
supersymmetric Ward identities to amplitudes of massive
scalars~\cite{Schwinn:2006ca}.  For this purpose, QCD with a massive
quark was embedded in a supersymmetric Yang-Mills theory with gluons
and gluinos, massive quarks $Q$ and two complex massive scalars
$\phi^+$ and $\phi^-$.  The external quark states defined by the
spinors~\eqref{eq:massive-spinors} are related to the scalars by SUSY
transformations parameterized by an anti-commuting spinor
$\kappa$
  \begin{equation}
\label{eq:susy}
\begin{aligned}
   \delta_\kappa \bar{Q}^\pm = \Gamma^\pm_\kappa(k) \bar{\phi}^\pm - \Sigma^\mp_\kappa(k,q) \bar{\phi}^\mp,
 & &
 \delta_\kappa \bar{\phi}^\pm = \Gamma^\mp_\kappa(k) \bar{Q}^\pm + \Sigma^\mp_\kappa(k,q) \bar{Q}^\mp,
 \nonumber \\
  \delta_\kappa Q^\pm =  -\Gamma^\pm_\kappa(k) \phi^\pm + \Sigma^\mp_\kappa(k,q) \phi^\mp,
 & &
 \delta_\kappa \phi^\pm = -\Gamma^\mp_\kappa(k) Q^\pm - \Sigma^\mp_\kappa(k,q) Q^\mp.
\end{aligned}
  \end{equation}
where
\begin{equation}
\label{eq:def-gamma}
  \Gamma^\pm_\kappa(k) =\sqrt 2\braket{\kappa\pm| k\mp},
 \;\;\;
  \Sigma_\kappa^\pm(k,q)=\sqrt 2 
  m\frac{\braket{q\pm|\kappa\mp}}{\braket{q\pm|k\mp}}\,.
\end{equation}
For the choice $\ket{\kappa+}\propto \theta\ket{q+}$, with a Grassmann number $\theta$, the terms proportional to the mass
drop out and the transformations are identical to those of massless particles.

For amplitudes with only positive helicity gluons one
finds the identity~\cite{Schwinn:2006ca}
\begin{equation}
\label{eq:master-swi}
\braket{\kappa 1} A(\bar Q_1^+,g_2^+,\dots,Q_n^-)
+m\frac{\braket{q\kappa}}{\braket{q1}}
 A(\bar Q_1^-,g_2^+,\dots,Q_n^-)
=\braket{\kappa n}  A(\bar \phi_1^+,g_2^+,\dots,\phi_n^-)\,.
\end{equation}
Amplitudes with gluinos generated by SUSY transformations of the gluons
can be shown to vanish at tree level.
Setting $\ket{\kappa+}\propto\theta\ket{q+}$
 or  $\ket{\kappa+}\propto\theta\ket{1+}$ one of the helicity combinations
on the left-hand side can be eliminated. In this way one finds the identities
\begin{align}
\label{eq:conserve+++}
  A(\bar Q_1^+,g_2^+,\dots,Q_n^-)&=\frac{\braket{nq}}{\braket{1q}}
    A(\bar \phi_1^+,g_2^+,\dots,\phi_n^-)\,,\\
A(\bar Q_1^-,g_2^+,\dots, Q_n^-)
&=\frac{\braket{1n}}{m} A(\bar \phi_1^+,g_2^+\dots,\phi_n^-)\,.
\label{eq:flip+++}
\end{align}
In a similar way one gets
\begin{equation}
    A(\bar Q_1^-,g_2^+,\dots,Q_n^+)=-\frac{\braket{1q}}{\braket{nq}}
    A(\bar \phi_1^-,g_2^+,\dots,\phi_n^+)\,. \label{eq:swi2}
\end{equation}
The amplitudes with a single negative helicity gluon  satisfy simple
SUSY-WIs provided the reference spinor is fixed in terms
of the momentum of the negative helicity gluon, $\ket{q+}=\ket{j+}$,
\begin{align}
\label{eq:neg-swi1}
\left.A_n(\bar{Q}_1^+,g_2^+,\dots,g_j^-,\dots, Q_n^+)\right|_{\ket{q+}=\ket{j+}}&=0,\\
\left.A_n(\bar{Q}_1^+,g_2^+,\dots,g_j^-,\dots, Q_n^-)\right|_{\ket{q+}=\ket{j+}}&=
\frac{\braket{nj}}{\braket{1j}}  
A_n(\bar{\phi}^+_1,g_2^+,\dots,g_j^-,\dots, \phi_n^-),
\label{eq:neg-swi2}\\
\left.A_n(\bar{Q}_1^-,g_2^+,\dots,g_j^-,\dots, Q_n^+)\right|_{\ket{q+}=\ket{j+}}&
=-\frac{\braket{1j}}{\braket{nj}}  
A_n(\bar{\phi}_1^-,g_2^+,\dots,g_j^-,\dots, \phi_n^+).\label{eq:neg-swi3}
\end{align}
In these identities $Q$ denotes an outgoing anti-quark. 
For the conventions used in the rules given in~\ref{sec:rules} there
is a sign change in the  identities~\eqref{eq:swi2} and~\eqref{eq:neg-swi3}

\section{Derivation of the helicity flip vertices}
\label{app:details}

In this appendix the details of the derivation of the
helicity flip vertex~\eqref{eq:flip--trafo} and~\eqref{eq:vertex+-+} will 
be filled in.

The helicity flip vertex with two negative helicity quarks is given in
terms of the redefined fields in~\eqref{eq:flip--trafo}. Inserting
the explicit form of the field redefinitions, this becomes
\begin{equation}
\label{eq:flip--def}
\mathcal{V}_{\bar \psi^-_1,B_{2},\dots, B_{n-1},\psi^-_n}
=-m (\sqrt 2 g)^{n-2}\frac{\braket{\eta 1}^2\braket{\eta n}^2}{
\braket{12}\dots\braket{(n-1)n}}\mathcal{S}_{\bar \psi^-_1,B_{2},\dots
B_{n-1},\psi^-_n}
\end{equation}
where the following two-fold sum was introduced:
\begin{equation}
\mathcal{S}_{\bar \psi^-_1,B_{2},\dots
B_{n-1},\psi^-_n} = 
\sum_{i=1}^{n-2}\sum_{j=i+1}^{n-1}
\left[  
\left(\frac{1}{ k_{(j+1,n),+}}
+\frac{1}{k_{(1,i),+}}\right)
\frac{(k_{(i+1,j),+})\braket{i(i+1)}\braket{j(j+1)}}{\braket{\eta i}
\braket{\eta (i+1)}\braket{j\eta} \braket{\eta (j+1)}  }
\right]\,.
\end{equation}

To perform the double sum, write $k_{(i+1,j)}=-(k_{1,i}+k_{j+1,n})$
in the first term. 
Applying the identity~\cite{Boels:2008ef} 
\begin{equation}
\label{eq:eikonal-b}
  \sum_{i=j}^{n-1}\frac{(k_{j,i})_+
\braket{i (i+1)}}{\braket{i\eta}\braket{\eta(i+1)}}
= \frac{\braket{\eta +|\fmslash k_{j,n-1}|n+}}{2\braket{\eta n}} 
\end{equation}
and the  so called eikonal identity 
(see e.g.~\cite{Mangano:1990by})
one of the sums in each of the terms in the resulting expression 
can be performed 
leading to
\begin{equation}
\begin{aligned}
  \mathcal{S}_{\bar \psi^-_1,B_{2},\dots B_{n-1},\psi^-_n}=&
\sum_{i=1}^{n-2}
\frac{1}{k_{(1,i),+}}
\frac{\braket{\eta+|\fmslash k_{i+1,n}|n+}\braket{i(i+1)}}{2 \braket{\eta i}
\braket{\eta (i+1)}\braket{\eta n}}\\
&+\sum_{j=2}^{n-1}
\frac{\braket{j(j+1)}}{\braket{\eta (j+1)}\braket{\eta j}}
\left(-\frac{\braket{\eta+|\fmslash k_{1,j}|j+}}{2k_{(j+1,n),+}\braket{\eta j}}+
\frac{\braket{1j}}{ \braket{\eta 1}\braket{\eta j}}\right)\,.
\end{aligned}
\end{equation}
Using momentum conservation to introduce a common factor of $\fmslash k_{1,i}$,
the terms with $i>2$ in the first sum 
and the first term in the second sum (up to $i=n-2$) can be combined using
the Schouten identity. The result involves $\braket{\eta+|\fmslash
  k_{1,i}|\eta+}=2k_{(1,i)+}$ in the numerator that cancels against the denominator.
Then everything is of the right shape to be combined with 
the remaining terms of the $j$-sum and the $i=1$ term
by another application of the Schouten identity. 
The final
sum in the resulting expression can be performed 
using the  eikonal identity:
\begin{equation}
\begin{aligned}
  \mathcal{S}_{\bar \psi^-_1,B_{2},\dots B_{n-1},\psi^-_n}=&
\frac{\braket{1n}}{\braket{\eta n}\braket{\eta 1} }
\sum_{i=1}^{n-1}\frac{\braket{i(i+1)}}{\braket{\eta i}
  \braket{\eta (i+1)}}
=\frac{\braket{1n}^2}{\braket{\eta n}^2\braket{\eta 1}^2 }\,.
\end{aligned}
\end{equation}
Inserting this expression into~\eqref{eq:flip--def} one obtains
 the result for the vertex function announced in~\eqref{eq:csw-m-}.

The helicity flip vertex with two positive helicity
quarks~\eqref{eq:vertex+-+} is derived in a similar way. 
Factoring out the factors common to all terms, the vertex function can 
be written as
\begin{equation}
\label{eq:def-S++}
  \mathcal{V}_{\bar \psi^+_1,B_{2},\dots ,\bar B_s,\dots B_{n-1},\psi^+_n}
\equiv
 m(\sqrt 2 g)^{n-1}
\frac{\braket{\eta s}^4}{
\braket{12}\dots\braket{(n-1)n}}
\mathcal{S}_{\bar \psi^+_1,B_{2},\dots,\bar B_s, \dots,B_{n-1},\psi^+_n}\,.
\end{equation}
In this case the evaluation of the double sum contained in the
quantity $\mathcal{S}_{\bar \psi^+_1,B_{2},\dots,\bar B_s,
  \dots,B_{n-1},\psi^+_n}$ is simpler since all explicit occurrences of
$+$ components of momenta can be eliminated using momentum
conservation and the sums can be evaluated by two subsequent
applications of the eikonal identity:
\begin{equation}
\begin{aligned}
\mathcal{S}_{\bar \psi^+_1,B_{2},\dots,\bar B_s,\dots,\psi^+_n}  &= 
-\sum_{i=1}^{s-1}\sum_{j=s}^{n-1}
\left[  
\left(\frac{1}{ k_{(j+1,n),+}}
+\frac{1}{k_{(1,i),+}}\right)
\frac{(k_{(1,i),+})(k_{(j+1,n),+})\braket{i(i+1)}\braket{j(j+1)}}{(k_{(i+1,j),+})
 \braket{i \eta}
\braket{\eta (i+1)}\braket{j\eta}\braket{\eta (j+1)}}
\right]\\
&=
\sum_{i=1}^{s-1}\sum_{j=s}^{n-1}
\left[ \frac{\braket{i(i+1)}\braket{j(j+1)}}{
 \braket{i \eta }
\braket{\eta (i+1)}\braket{j\eta}\braket{\eta (j+1)} }
\right]
=\frac{\braket{1s}\braket{sn}}{
 \braket{\eta 1}
\braket{\eta s}^2\braket{\eta n}}\,.
\end{aligned}
\end{equation}
Inserting this into~\eqref{eq:def-S++}
 one obtains the result given in~\eqref{eq:vertex+-+}.
\section{Four- and five point amplitudes with massive quarks}
\label{app:amplitudes}
In order to compare to results from the CSW rules we collect
some results for four and five point amplitudes with massive quarks.
All results follow the conventions discussed in section~\ref{sec:rules}, i.e.
 $Q$ denotes an incoming quark.
From a Feynman diagram calculation one obtains the four point amplitude with a massive quark pair and one negative helicity gluon as
\begin{equation}
\label{eq:feyn4}
  A(\bar Q_1^{\sigma_1},g_2^-,g_3^+,Q_4^{\sigma_4})=
\ii \frac{\braket{3+|\fmslash k_1|2+}}{
2(k_1\cdot k_2)(k_2\cdot k_3)}
\bar u(k_1,\sigma_1)
\left( \ket{2+}\bra{3+}+\ket{3-}\bra{2-}\right)u(k_4,\sigma_4)\,.
\end{equation}
Using on-shell recursion relations for `stripped amplitudes' with 
the quark polarization spinors removed~\cite{Badger:2005jv} one finds for the 2-quark three gluon amplitudes with a negative
helicity gluon adjacent to a massive quark
\begin{multline}
\label{eq:npt-++}
 A_5(\bar Q_1^{\sigma_1},g_2^-,g_{3}^+,g_{4}^+,Q_5^{\sigma_5})=\\
\frac{2^{3/2}\ii}{ 
\braket{4-|\fmslash k_{2,3}\fmslash k_1|2+}}
\Biggl[
 \frac{\braket{2-|\fmslash k_{5}\fmslash k_1|2+}}{\braket{23}\braket{34}
k_{2,4}^2 }
\bar u(k_1,\sigma_1)\bigl(
\fmslash k_{3,4}\ket{2+}\bra{2-}-\ket{2+}\bra{2-}\fmslash k_{3,4}
\bigr))u(k_5,\sigma_5)
\\
-
\frac{m
 \braket{3+|\fmslash k_1|2+ }
\sbraket{34}}{ y_{1,2}y_{1,3}\sbraket{32} }
\bar u(k_1,\sigma_1)
\bigl( \ket{2+}\bra{3+}\fmslash k_{1,2}+m
\ket{3-}\bra{2-}\bigr)u(k_5,\sigma_5)
\Biggr]\,.
\end{multline}

We will also consider 
four-quark amplitudes with two massive quarks $Q$ 
and two massless quarks $q$. 
The results from 
 conventional Feynman rules are
\begin{align}
\label{eq:4q-flip}
  A_4(\bar Q_1^+,q_2^+,\bar{q}_3{}^-,Q_4^+)
 =& 
 \frac{2i m \braket{ \eta 3 }\braket{2+|\fmslash k_3 | \eta+ }}{
   \braket{1 \eta} \braket{\eta 4}(k_{2}-k_3)^2} 
 =\frac{2i m \braket{ \eta 3 }^2}{
   \braket{1 \eta} \braket{4\eta }\braket{23}} \,, \\
 A_4(\bar Q_1^+,q_2^-,\bar{q}_3{}^+,Q_4^+)
 =& \frac{2i m \braket{ \eta 2 }^2}{
   \braket{1 \eta} \braket{4\eta }\braket{23}} \\
   A_4(\bar Q_1^+,q_2^+,\bar{q}_3{}^-,Q_4^-)\,,
 =& 
 \frac{-2i}{\braket{1\eta}\sbraket{\eta 4} 
(k_2-k_3)^2}
 \Bigl(
   \braket{2+|\fmslash k_1 | \eta+ }
\braket{ \eta+ | \fmslash k_4 |3+ } +m^2 \sbraket{ 2\eta }
\braket{ \eta 3}
 \Bigr)\,.
\label{eq:4q-conserve}
\end{align}

\providecommand{\href}[2]{#2}\begingroup\raggedright\endgroup



\begin{thebibliography}{10}

\bibitem{Cachazo:2004kj}
F.~Cachazo, P.~Svr\v{c}ek, and E.~Witten {\em JHEP} {\bf 09} (2004)  006,
\href{http://arxiv.org/abs/hep-th/0403047}{{\tt hep-th/0403047}}\relax
\relax
\bibitem{Parke:1986gb}
S.~J. Parke and T.~R. Taylor
{\em Phys. Rev. Lett.} {\bf 56} (1986)  2459\relax
\relax
\bibitem{Berends:1987me}
F.~A. Berends and W.~T. Giele
{\em Nucl. Phys.} {\bf B306} (1988)  759\relax
\relax
\bibitem{Wu:2004fb}
J.-B. Wu and C.-J. Zhu {\em JHEP} {\bf 07} (2004)  032,
\href{http://arxiv.org/abs/hep-th/0406085}{{\tt hep-th/0406085}}; \relax
\relax
J.-B. Wu and C.-J. Zhu {\em JHEP} {\bf 09} (2004)  063,
\href{http://arxiv.org/abs/hep-th/0406146}{{\tt hep-th/0406146}}\relax
\relax
\bibitem{Georgiou:2004wu}
G.~Georgiou and V.~V. Khoze {\em JHEP} {\bf 05} (2004)  070,
\href{http://arxiv.org/abs/hep-th/0404072}{{\tt hep-th/0404072}}; \relax
\relax
G.~Georgiou, E.~W.~N. Glover, and V.~V. Khoze {\em JHEP} {\bf 07} (2004)  048,
\href{http://arxiv.org/abs/hep-th/0407027}{{\tt hep-th/0407027}}\relax
\relax
\bibitem{Britto:2005fq}
R.~Britto, F.~Cachazo, B.~Feng, and E.~Witten {\em Phys. Rev. Lett.} {\bf 94}
  (2005)  181602,
\href{http://arxiv.org/abs/hep-th/0501052}{{\tt hep-th/0501052}}\relax
\relax
\bibitem{Dinsdale:2006sq}
M.~Dinsdale, M.~Ternick, and S.~Weinzierl {\em JHEP} {\bf 03} (2006)  056,
  \href{http://arxiv.org/abs/hep-ph/0602204}{{\tt hep-ph/0602204}}; \relax
\relax
C.~Duhr, S.~H\"oche, and F.~Maltoni {\em JHEP} {\bf 08} (2006)  062,
  \href{http://arxiv.org/abs/hep-ph/0607057}{{\tt hep-ph/0607057}}; \relax
\relax
 T.~Gleisberg, S.~H\"oche, F.~Krauss and R.~Matyszkiewicz,
  {\tt arXiv:0808.3672 [hep-ph]}.
\bibitem{Brandhuber:2004yw}
A.~Brandhuber, B.~J. Spence, and G.~Travaglini {\em Nucl. Phys.} {\bf B706}
  (2005)  150--180,
\href{http://arxiv.org/abs/hep-th/0407214}{{\tt hep-th/0407214}}; \relax
\relax
J.~Bedford, A.~Brandhuber, B.~Spence, and G.~Travaglini {\em Nucl. Phys.} {\bf
  B706} (2005)  100--126,
\href{http://arxiv.org/abs/hep-th/0410280}{{\tt hep-th/0410280}}; \relax
\relax
C.~Quigley and M.~Rozali {\em JHEP} {\bf 01} (2005)  053,
\href{http://arxiv.org/abs/hep-th/0410278}{{\tt hep-th/0410278}}; \relax
\relax
E.~W.~N. Glover, V.~V. Khoze, and C.~Williams
  \href{http://dx.doi.org/10.1088/1126-6708/2008/08/033}{{\em JHEP} {\bf 08}
  (2008)  033},
\href{http://arxiv.org/abs/0805.4190}{{\tt arXiv:0805.4190 [hep-th]}}\relax
\relax
\bibitem{Bedford:2004nh}
J.~Bedford, A.~Brandhuber, B.~J. Spence, and G.~Travaglini {\em Nucl. Phys.}
  {\bf B712} (2005)  59--85,
\href{http://arxiv.org/abs/hep-th/0412108}{{\tt hep-th/0412108}}; \relax
\relax
A.~Brandhuber, B.~Spence, G.~Travaglini, and K.~Zoubos
  \href{http://dx.doi.org/10.1088/1126-6708/2007/07/002}{{\em JHEP} {\bf 07}
  (2007)  002},
\href{http://arxiv.org/abs/0704.0245}{{\tt arXiv:0704.0245 [hep-th]}}\relax
\relax
\bibitem{Ettle:2007qc}
J.~H. Ettle, C.-H. Fu, J.~P. Fudger, P.~R.~W. Mansfield, and T.~R. Morris {\em
  JHEP} {\bf 05} (2007)  011,
\href{http://arxiv.org/abs/hep-th/0703286}{{\tt hep-th/0703286}}\relax
\relax
\bibitem{Birthwright:2005ak}
T.~G. Birthwright, E.~W.~N. Glover, V.~V. Khoze, and P.~Marquard {\em JHEP}
  {\bf 05} (2005)  013,
\href{http://arxiv.org/abs/hep-ph/0503063}{{\tt hep-ph/0503063}}; \relax
\relax
T.~G. Birthwright, E.~W.~N. Glover, V.~V. Khoze, and P.~Marquard {\em JHEP}
  {\bf 07} (2005)  068,
\href{http://arxiv.org/abs/hep-ph/0505219}{{\tt hep-ph/0505219}}; \relax
\relax
C.~Duhr and F.~Maltoni
\href{http://arxiv.org/abs/0808.3319}{{\tt arXiv:0808.3319 [hep-ph]}}\relax
\relax
\bibitem{Witten:2003nn}
E.~Witten {\em Commun. Math. Phys.} {\bf 252} (2004)  189--258,
\href{http://arxiv.org/abs/hep-th/0312171}{{\tt hep-th/0312171}}\relax
\relax
\bibitem{Bern:2004ba}
Z.~Bern, D.~Forde, D.~A. Kosower, and P.~Mastrolia {\em Phys. Rev.} {\bf D72}
  (2005)  025006,
\href{http://arxiv.org/abs/hep-ph/0412167}{{\tt hep-ph/0412167}}\relax
\relax
\bibitem{Dixon:2004za}
L.~J. Dixon, E.~W.~N. Glover, and V.~V. Khoze {\em JHEP} {\bf 12} (2004)  015,
\href{http://arxiv.org/abs/hep-th/0411092}{{\tt hep-th/0411092}}; \relax
\relax
S.~D. Badger, E.~W.~N. Glover, and V.~V. Khoze {\em JHEP} {\bf 03} (2005)  023,
\href{http://arxiv.org/abs/hep-th/0412275}{{\tt hep-th/0412275}}; \relax
\relax
S.~D. Badger, E.~W.~N. Glover, and K.~Risager
  \href{http://dx.doi.org/10.1088/1126-6708/2007/07/066}{{\em JHEP} {\bf 07}
  (2007)  066},
\href{http://arxiv.org/abs/0704.3914}{{\tt arXiv:0704.3914 [hep-ph]}}; \relax
\relax
E.~W.~N. Glover, P.~Mastrolia, and C.~Williams
  \href{http://dx.doi.org/10.1088/1126-6708/2008/08/017}{{\em JHEP} {\bf 08}
  (2008)  017},
\href{http://arxiv.org/abs/0804.4149}{{\tt arXiv:0804.4149 [hep-ph]}}\relax
\relax
\bibitem{Risager:2005vk}
K.~Risager {\em JHEP} {\bf 12} (2005)  003,
\href{http://arxiv.org/abs/hep-th/0508206}{{\tt hep-th/0508206}}\relax
\relax
\bibitem{Gorsky:2005sf}
A.~Gorsky and A.~Rosly {\em JHEP} {\bf 01} (2006)  101,
\href{http://arxiv.org/abs/hep-th/0510111}{{\tt hep-th/0510111}}\relax
\relax
\bibitem{Mansfield:2005yd}
P.~Mansfield {\em JHEP} {\bf 03} (2006)  037,
\href{http://arxiv.org/abs/hep-th/0511264}{{\tt hep-th/0511264}}\relax
\relax
\bibitem{Ettle:2006bw}
J.~H. Ettle and T.~R. Morris {\em JHEP} {\bf 08} (2006)  003,
\href{http://arxiv.org/abs/hep-th/0605121}{{\tt hep-th/0605121}}\relax
\relax
\bibitem{Boels:2007qn}
R.~Boels, L.~Mason, and D.~Skinner {\em Phys. Lett.} {\bf B648} (2007)  90--96,
\href{http://arxiv.org/abs/hep-th/0702035}{{\tt hep-th/0702035}}; \relax
{\it For a review see} W.~Jiang \href{http://arxiv.org/abs/0809.0328}{{\tt arXiv:0809.0328 [hep-th]}}.
{D.Phil. thesis, University of Oxford}\relax
\relax
\relax
\bibitem{Mason:2005zm}
L.~J. Mason {\em JHEP} {\bf 10} (2005)  009,
\href{http://arxiv.org/abs/hep-th/0507269}{{\tt hep-th/0507269}}; \relax
\relax
R.~Boels, L.~Mason, and D.~Skinner {\em JHEP} {\bf 02} (2007)  014,
\href{http://arxiv.org/abs/hep-th/0604040}{{\tt hep-th/0604040}} \relax
\relax
\bibitem{Boels:2007pj}
R.~Boels and C.~Schwinn
  \href{http://dx.doi.org/10.1016/j.physletb.2008.02.038}{{\em Phys. Lett.}
  {\bf B662} (2008)  80--86},
\href{http://arxiv.org/abs/0712.3409}{{\tt arXiv:0712.3409 [hep-th]}}\relax
\relax
\bibitem{Boels:2008ef}
R.~Boels and C.~Schwinn
  \href{http://dx.doi.org/10.1088/1126-6708/2008/07/007}{{\em JHEP} {\bf 07}
  (2008)  007},
\href{http://arxiv.org/abs/0805.1197}{{\tt arXiv:0805.1197 [hep-th]}}\relax
\relax
\bibitem{Badger:2005zh}
S.~D. Badger, E.~W.~N. Glover, V.~V. Khoze, and P.~Svr\v{c}ek {\em JHEP} {\bf 07}
  (2005)  025,
\href{http://arxiv.org/abs/hep-th/0504159}{{\tt hep-th/0504159}}; \relax
\relax
D.~Forde and D.~A. Kosower {\em Phys. Rev.} {\bf D73} (2006)  065007,
\href{http://arxiv.org/abs/hep-th/0507292}{{\tt hep-th/0507292}}; \relax
\relax
P.~Ferrario, G.~Rodrigo, and P.~Talavera {\em Phys. Rev. Lett.} {\bf 96} (2006)
   182001,
\href{http://arxiv.org/abs/hep-th/0602043}{{\tt hep-th/0602043}}\relax
\relax
\bibitem{Schwinn:2007ee}
C.~Schwinn and S.~Weinzierl {\em JHEP} {\bf 04} (2007)  072,
\href{http://arxiv.org/abs/hep-ph/0703021}{{\tt hep-ph/0703021}}\relax
\relax
\bibitem{Ettle:2008ey}
J.~H. Ettle, T.~R. Morris, and Z.~Xiao
{\em JHEP} {\bf 08} (2008) 103,
\href{http://arxiv.org/abs/0805.0239}{{\tt arXiv:0805.0239 [hep-th]}}\relax
\relax
\bibitem{Boels:2008du}
R.~Boels and C.~Schwinn \href{http://arxiv.org/abs/0805.4577}{{\tt
  arXiv:0805.4577 [hep-th]}}.
To appear in the proceedings of the 9th Workshop On Elementary Particle Theory:
  Loops And Legs In Quantum Field Theory, 20-25 Apr 2008, Sondershausen,
  Germany\relax
\relax
\bibitem{Grisaru:1976vm}
M.~T. Grisaru, H.~N. Pendleton, and P.~van Nieuwenhuizen
{\em Phys. Rev.} {\bf D15} (1977)  996; \relax
\relax
M.~T. Grisaru and H.~N. Pendleton
{\em Nucl. Phys.} {\bf B124} (1977)  81;\relax
\relax
S.~J. Parke and T.~R. Taylor
{\em Phys. Lett.} {\bf B157} (1985)  81; \relax
\relax
Z.~Kunszt
{\em Nucl. Phys.} {\bf B271} (1986)  333\relax
\relax
\bibitem{Schwinn:2006ca}
C.~Schwinn and S.~Weinzierl {\em JHEP} {\bf 03} (2006)  030,
\href{http://arxiv.org/abs/hep-th/0602012}{{\tt hep-th/0602012}}\relax
\relax
\bibitem{Schwinn:2005pi}
C.~Schwinn and S.~Weinzierl {\em JHEP} {\bf 05} (2005)  006,
\href{http://arxiv.org/abs/hep-th/0503015}{{\tt hep-th/0503015}}\relax
\relax
\bibitem{Bauer:2000yr}
C.~W. Bauer, S.~Fleming, D.~Pirjol, and I.~W. Stewart
  \href{http://dx.doi.org/10.1103/PhysRevD.63.114020}{{\em Phys. Rev.} {\bf
  D63} (2001)  114020},
\href{http://arxiv.org/abs/hep-ph/0011336}{{\tt arXiv:hep-ph/0011336}}; \relax
\relax
C.~W. Bauer, D.~Pirjol, and I.~W. Stewart
  \href{http://dx.doi.org/10.1103/PhysRevD.65.054022}{{\em Phys. Rev.} {\bf
  D65} (2002)  054022},
\href{http://arxiv.org/abs/hep-ph/0109045}{{\tt arXiv:hep-ph/0109045}}; \relax
\relax
M.~Beneke, A.~P. Chapovsky, M.~Diehl, and T.~Feldmann {\em Nucl. Phys.} {\bf
  B643} (2002)  431--476,
\href{http://arxiv.org/abs/hep-ph/0206152}{{\tt hep-ph/0206152}}\relax
\relax
\bibitem{Rothstein:2003wh}
I.~Z. Rothstein \href{http://dx.doi.org/10.1103/PhysRevD.70.054024}{{\em Phys.
  Rev.} {\bf D70} (2004)  054024},
\href{http://arxiv.org/abs/hep-ph/0301240}{{\tt arXiv:hep-ph/0301240}}; \relax
\relax
A.~K. Leibovich, Z.~Ligeti, and M.~B. Wise
  \href{http://dx.doi.org/10.1016/S0370-2693(03)00565-3}{{\em Phys. Lett.} {\bf
  B564} (2003)  231--234},
\href{http://arxiv.org/abs/hep-ph/0303099}{{\tt arXiv:hep-ph/0303099}}; \relax
\relax
H.~Boos, T.~Feldmann, T.~Mannel, and B.~D. Pecjak
  \href{http://dx.doi.org/10.1103/PhysRevD.73.036003}{{\em Phys. Rev.} {\bf
  D73} (2006)  036003},
\href{http://arxiv.org/abs/hep-ph/0504005}{{\tt arXiv:hep-ph/0504005}}\relax
\relax
\bibitem{Chalmers:1998jb}
G.~Chalmers and W.~Siegel {\em Phys. Rev.} {\bf D59} (1999)  045013,
\href{http://arxiv.org/abs/hep-ph/9801220}{{\tt hep-ph/9801220}}\relax
\relax
\bibitem{Kleiss:1985yh}
R.~Kleiss and W.~J. Stirling
{\em Nucl. Phys.} {\bf B262} (1985)  235--262; \relax
\relax
R.~Kleiss and W.~J. Stirling
\href{http://dx.doi.org/10.1016/0370-2693(86)90454-5}{{\em Phys. Lett.} {\bf
  B179} (1986)  159}; \relax
\relax
A.~Ballestrero and E.~Maina {\em Phys. Lett.} {\bf B350} (1995)  225--233,
\href{http://arxiv.org/abs/hep-ph/9403244}{{\tt hep-ph/9403244}}; \relax
\relax
S.~Dittmaier {\em Phys. Rev.} {\bf D59} (1998)  016007,
\href{http://arxiv.org/abs/hep-ph/9805445}{{\tt hep-ph/9805445}}; \relax
\relax
J.~van~der Heide, E.~Laenen, L.~Phaf, and S.~Weinzierl {\em Phys. Rev.} {\bf
  D62} (2000)  074025,
\href{http://arxiv.org/abs/hep-ph/0003318}{{\tt hep-ph/0003318}}\relax
\bibitem{Feng:2006yy}
H.~Feng and Y.-t. Huang
\href{http://arxiv.org/abs/hep-th/0611164}{{\tt hep-th/0611164}}\relax
\relax
\bibitem{Mangano:1990by}
M.~L. Mangano and S.~J. Parke {\em Phys. Rept.} {\bf 200} (1991)  301--367,
\href{http://arxiv.org/abs/hep-th/0509223}{{\tt hep-th/0509223}}; \relax
\relax
L.~J. Dixon, {\it Calculating scattering amplitudes efficiently},  in {\em
  {QCD} and beyond: Proceedings of TASI 95}, D.~Soper, ed., pp.~539--584.
\newblock 1996.
\newblock
\href{http://arxiv.org/abs/hep-ph/9601359}{{\tt hep-ph/9601359}}\relax
\relax
\bibitem{Kosower:2004yz}
D.~A. Kosower {\em Phys. Rev.} {\bf D71} (2005)  045007,
\href{http://arxiv.org/abs/hep-th/0406175}{{\tt hep-th/0406175}}\relax
\relax
\bibitem{Boels:private}
R.~Boels. Private communication\relax
\relax
\bibitem{Badger:2005jv}
S.~D. Badger, E.~W.~N. Glover, and V.~V. Khoze {\em JHEP} {\bf 01} (2006)  066,
\href{http://arxiv.org/abs/hep-th/0507161}{{\tt hep-th/0507161}}; \relax
\relax
K.~J. Ozeren and W.~J. Stirling {\em Eur. Phys. J.} {\bf C48} (2006)  159--168,
\href{http://arxiv.org/abs/hep-ph/0603071}{{\tt hep-ph/0603071}}; \relax
\relax
A.~Hall \href{http://dx.doi.org/10.1103/PhysRevD.77.025011}{{\em Phys. Rev.}
  {\bf D77} (2008)  025011},
\href{http://arxiv.org/abs/0710.1300}{{\tt arXiv:0710.1300 [hep-ph]}}\relax
\relax
\end{thebibliography}
\end{document}